\documentclass[fleqn,usenatbib]{mnras}

\usepackage{newtxtext,newtxmath}

\usepackage[T1]{fontenc}
\usepackage{ae,aecompl}

\usepackage{graphicx}	
\usepackage{amsmath}	

\newcommand{\diff}{\mathrm{d}}
\newcommand{\sh}{\mathrm{sh}}
\newcommand{\view}{\theta_\mathrm{v}}

\newcommand{\emisd}{\epsilon' _{\nu '}}
\newcommand{\numd}{\nu'_\mathrm{m}}
\newcommand{\nucd}{\nu'_\mathrm{c}}
\newcommand{\me}{m_\mathrm{e}}
\newcommand{\mpr}{m_\mathrm{p}}
\newcommand{\qe}{q_\mathrm{e}}
\newcommand{\eB}{\varepsilon_\mathrm{B}}
\newcommand{\ee}{\varepsilon_\mathrm{e}}
\newcommand{\BM}{\mathrm{BM}}
\newcommand{\ST}{\mathrm{ST}}
\newcommand{\simpropto}{\mathrel{\vcenter{
			\offinterlineskip\halign{\hfil$##$\cr
				\propto\cr\noalign{\kern2pt}\sim\cr\noalign{\kern-2pt}}}}}
\defcitealias{TI20}{TI20}
\defcitealias{TI21}{TI21}
\defcitealias{KW05}{KW05}
\newcommand{\Juttner}{J\"{u}ttner}
\newcommand{\betau}{\beta_\mathrm{u}}
\newcommand{\betad}{\beta_\mathrm{d}}

\title[Probing particle acceleration with GRB afterglows]{Probing particle acceleration at trans-relativistic shocks with off-axis gamma-ray burst afterglows}

\author[Takahashi et al.]{
Kazuya Takahashi$^{1,2}$\thanks{E-mail: kazuya.takahashi@yukawa.kyoto-u.ac.jp}, 
Kunihito Ioka$^{2}$,
Yutaka Ohira$^{3}$,
and Hendrik J.~van Eerten$^{4}$
\\
$^{1}$Research Center for the Early Universe, The University of Tokyo, 7-3-1 Hongo, Bunkyo-ku, Tokyo 113-0033, Japan\\
$^{2}$Center for Gravitational Physics, Yukawa Institute for Theoretical Physics, Kyoto University, Kyoto, 606-8502, Japan\\
$^{3}$Department of Earth and Planetary Science, The University of Tokyo, 7-3-1 Hongo, Bunkyo-ku, Tokyo 113-0033, Japan\\
$^{4}$Department of Physics, University of Bath, Claverton Down, Bath BA2 7AY, United Kingdom
}

\date{Accepted XXX. Received YYY; in original form ZZZ}

\pubyear{2021}

\begin{document}
\label{firstpage}
\pagerange{\pageref{firstpage}--\pageref{lastpage}}
\maketitle

\begin{abstract}
Particle acceleration is expected to be different between relativistic and non-relativistic collisionless shocks. We show that electromagnetic counterparts to gravitational waves (GWs), gamma-ray burst (GRB) afterglows, are ideal targets for observing trans-relativistic evolution of accelerated electron distribution because the GWs spot nearby GRBs with off-axis jets, otherwise missed in gamma-ray observations. 
We find that the relativistic spectral slope begins to change steeply near the peak time of the light curve and approaches the non-relativistic limit in about 10 times the peak time.
The trans-relativistic evolution of the afterglow synchrotron spectrum is consistent with GRB~170817A observations within errors, and will be measurable in similar but more distant events at a GW horizon $\sim 200$~Mpc in a denser environment.
We roughly estimate that such events represent a fraction of 10-50 per cent of the GRB~170817A-like off-axis short GRBs.
We also find that the spectral evolution does not depend on the jet structure if their light curves are similar to each other.
\end{abstract}




\section{Introduction}
Particle acceleration across collisionless shock waves, called diffusive shock acceleration or first-order Fermi acceleration, is thought to be a mechanism to produce non-thermal high-energy particles in astrophysical phenomena \citep[][for reviews]{Drury83,BE,Axford94,SKL15,UAO19,Marcowith20}.
While the details of the acceleration mechanism are still under debate,
particle acceleration appears to work both in relativistic and non-relativistic shocks as indicated by observations of gamma-ray bursts (GRBs) and supernova remnants (SNRs).
In relativistic flow, the energy spectral index, or the power-law index, of the accelerated particles can be different from that in the non-relativistic case, since the distribution of the momentum measured near the shock front at the shock rest frame is far from isotropic due to more particles moving downstream than upstream \citep{KS87, HD88, Kirk00, KW05, Lavi20}.
However, it has not been confirmed in observations that the power-law index changes with shock speed through the relativistic to non-relativistic regimes, 
whereas the evolution of the index in non-relativistic shocks has been recognized by comparing SNRs with different ages \citep{G85,Bell11,Bozzetto17}.

Trans-relativistic shocks driven by high-velocity ejecta of some peculiar supernovae are expected to accelerate ultra-high-energy cosmic rays \citep{Wang07,Budnik08,Chakraborti11}.
However, these are less energetic than GRBs and occur in dense regions, causing the shock to quickly decelerate and become non-relativistic. Therefore, much remains unprobed about the full evolution of particle acceleration from trans-relativistic shocks.

GRB afterglows are a unique potential source that could be used to probe a mechanism of particle acceleration at trans-relativistic shocks. 
A GRB afterglow is produced by synchrotron radiation emitted by non-thermal electrons accelerated at a shock wave ahead of a relativistic collimated outflow (jet).
While producing afterglow emission, the shock wave slows down from a relativistic speed to a non-relativistic speed by sweeping up the ambient stellar medium.
The change of the electron power-law index $p$ is imprinted in the time evolution of the spectral slopes in GRB afterglows.
Suppose that the accelerated electrons have a power-law energy distribution $\propto \gamma_\mathrm{e} ^{\prime -p}$, where $\gamma_\mathrm{e}^\prime$ is the Lorentz factor of an electron.
Then, the logarithmic slope between the synchrotron characteristic frequency $\nu_\mathrm{m}$ and cooling frequency $\nu_\mathrm{c}$ is given by $-(p-1)/2$ in the slow cooling phase \citep{Sari98}.

GRB~170817A, the short GRB directly associated with the first gravitational wave signal GW170817 from a binary neutron star merger, is a great example presenting afterglow spectra with a beautiful single power-law extending from radio to X-ray \citep{PRL170817,170817multi,170817FermiGBM,170817Integral}. 
A successful launch of a relativistic jet in GRB~170817A is supported by the detection of superluminal motion of a compact radio source \citep{superluminal,Ghirlanda19}.
The jet was observed from off-axis (i.e., the jet axis and the line of sight was misaligned), leading to a faint gamma-ray emission \citep[][and references therein]{170817gamma,IN18,IN19,SG22} and early rising afterglow light curves \citep{Alexander17,Hallinan17,Margutti17,Troja17}.
The slow rising afterglow light curves also rejected a uniform top-hat jet \citep{Mooley18,Troja18a} and revealed a launch of a structured jet, while various jet structures can explain the observations (see, however, \citet{Lamb20} for an alternative scenario with a refreshed shock in an off-axis uniform jet).
Often assumed structured jets are a Gaussian jet or a power-law jet \citep{DAvanzo18,GG18,LK18,Lyman18,Resmi18,Ghirlanda19,rapiddeclineHST,Troja19,Ryan,Troja20}.
Other non-trivial jet structures such as hollow-cone jets and spindle jets are also possible candidates as recently discovered by \citet{TI20,TI21}, while the consistency of a hollow-cone jet was also confirmed by \citet{Nathanail20,Nathanail21}.
As predicted by \citet{Troja18a}, the rapid decline of the afterglow after the peak \citep{rapiddecline,rapiddeclineHST,Hajela20,Panchromatic,Troja20,Balas} revealed a distinctive signature of a successfully launched structured jet, setting it apart from a chocked jet scenario.
The observed electron power-law index derived from the afterglow spectrum falls within the range predicted in a theory of particle acceleration at trans-relativistic shock \citep{KW05} as mentioned by \citet{Margutti18}
and is consistent with being constant in time within observational errors \citep{DAvanzo18,Dobie18,Margutti18,Fong19,Kilpatrick}.

However, the constancy of the electron power-law index $p$ measured in the GRB~170817A afterglow is not so conclusive due to large observational errors.
It could be also a fortunate coincidence that the electron power-law index in GRB~170817A falls within the range predicted in theory, 
since the electron power-law index observed in other GRBs is not represented by a single universal value but spreads over a wide range \citep{PK02,Shen06,Starling08,Curran09,Curran10,Fong15,Gompertz18,Troja19}.
Thus, the value or time evolution of the electron power-law index in nature remains inconclusive.

This paper investigates a potential use of afterglows of off-axis short GRBs for probing particle acceleration at trans-relativistic shock.
Nearby short GRBs are expected to be off-axis and are an interesting source in the coming multi-messenger era, since they will be found as an electromagnetic counterpart of a gravitational wave signal from a binary neutron star merger as demonstrated by GRB~170817A associated with GW170817 \citep{PRL170817,170817multi,170817FermiGBM,170817Integral}.
An afterglow emission of an off-axis GRB can be found as a counterpart of a gravitational wave,
even if the prompt gamma-ray emission is faint and is not detected.
The afterglow can be more luminous than those in cosmological GRBs, depending on the distance, ambient number density, and viewing angle \citep[e.g.,][]{GNP19,TI21}, which facilitates the measurement of the electron power-law index.
We theoretically calculate afterglow spectra and the time evolution of the electron power-law index obtained from the spectra by using a particle acceleration model.
We demonstrate how particle acceleration at trans-relativistic shocks can be probed by comparing the theoretical prediction with observations.

The paper is organized as follows. Section~\ref{sec.method} introduces the method to calculate GRB afterglows. 
In Section~\ref{sec.results}, we present our numerical results showing how the electron power-law index obtained from the spectral slope in off-axis GRB afterglows evolves with time as the shock is decelerated and show that there would be a chance to probe particle acceleration by future observations of off-axis GRB afterglows. 
Section~\ref{sec.discussion} is devoted to discussion of the event rate and effects of sideways expansion of the jet.
We summarize our findings and draw our conclusions in Section~\ref{sec.conclusion}.
Throughout the paper, we attach a prime to quantities evaluated in the fluid rest frame.
Quantities without a prime are evaluated in the laboratory frame unless otherwise mentioned.

\section{Method}\label{sec.method}
We calculate afterglow light curves and spectra of off-axis GRBs and discuss the possibility to probe particle acceleration across trans-relativistic shocks.
The particle acceleration process is imprinted in GRB afterglow spectra through the energy power-law index of the electrons accelerated at the shock wave of a GRB jet \citep{Sari98}.
We focus on off-axis GRBs, since the detection number of the nearby off-axis events would increase in the era of the multi-messenger astronomy including gravitational waves.
We review a theoretical model for calculating off-axis GRB afterglows in Section~\ref{sec.method.afterglow}. In Section~\ref{sec.method.jet}, we introduce a non-uniform structured jet model to which the afterglow model is applied.

\subsection{Off-axis GRB afterglow model}\label{sec.method.afterglow}
We theoretically calculate an off-axis afterglow light curve and the spectrum for given jet structure and afterglow parameters. We apply the standard model of GRB afterglows by \citet{Sari98},
where the afterglow is produced by the synchrotron radiation emitted from the non-thermal electrons accelerated across the forward shock. In the standard model, the energy spectrum of the accelerated electrons is given by a single power law with an index $p$. Microscopic physics such as the amplification of magnetic fields and particle acceleration through the shock wave is modelled by introducing phenomenological parameters, $\eB$ and $\ee$, respectively. 

While recent work \citep{Ayache22} explores the implications of the trans-relativistic evolution of slope $p$ in the context of high-resolution jet simulations, their study is limited to on-axis light curves and jets starting from top-hat initial conditions. In this work, we for the first time combine trans-relativistic acceleration models with structured jet profiles and off-axis observations.
We take into account the dependence of $p$ on the shock Lorentz factor $\Gamma_\sh$ by applying a particle acceleration model, whereas the index $p$ is usually assumed as a constant in the literature.
We adopt the model of \citet{KW05}, for the sake of argument, as a first attempt.
The other basic picture of the model is essentially the same as in \citet{TI20,TI21} (hereafter \citetalias{TI20} and \citetalias{TI21}, respectively) but is partly revised as reviewed below.
\footnote{We caution that the way to calculate GRB afterglows is not clearly mentioned and is not traceable in many previous papers, although there are many options in equations, which prevents readers from quantitative comparison. For examples, the coefficients in Equations~\eqref{eq.emisdpeak}-\eqref{eq.nucd} are based on \citet{RL}, \citet{Granot99}, and \citet{Eerten10} and are different from those in \citet{Sari98}.
The care for the Newtonian regime in the non-thermal electron distribution given by Equations~\eqref{eq.gmin} and \eqref{eq.nR} is another example, which is necessary for calculating a non-relativistic shock \citep{vanEerten13} but is rarely mentioned in previous papers.}
First, Section~\ref{sec.method.dynamics} briefly reviews the dynamics of a forward shock in a GRB jet. Then, Section~\ref{sec.method.p} introduces a particle acceleration model and physical quantities in a shocked flow. Finally, Section~\ref{sec.method.obsFlux} shows equations for calculating afterglow fluxes.

\subsubsection{Evolution of the shock wave}\label{sec.method.dynamics}
The jet is assumed to be axi-symmetric and adiabatically propagates into a stationary, cold, uniform medium with a number density $n_0$. The jet has an angular structure where the isotropic equivalent energy is given as $E(\theta)$, where $\theta$ is the polar angle measured from the jet axis. We assume that each jet segment expands radially as if it were a portion of the isotropic blast wave with the same isotropic equivalent energy. This assumption holds well for a relativistic shock unless it is decelerated sufficiently below a local sound speed and each segment interacts with each other \citep{KG03,ZM09,EM12} (we separately consider the effect of lateral expansion of a jet in Sec.~\ref{sec.sideways}). Then, the dynamics of each segment of the shock wave is described by a self-similar solution of \citet{BM}. As the shock slows down to a non-relativistic speed by sweeping up the ambient matter, the shock dynamics is better described by the Sedov-Taylor solution \citep{Sedov,Taylor}. Hence, we smoothly connect the Blandford-McKee (BM) and Sedov-Taylor (ST) solutions by giving the shock dynamics as follows \citep{Eerten10}:
\begin{align}
\label{eq.BM-STsh}
\Gamma_\sh^2 \beta_\sh^2 &= \mathrm{min}(C_\BM^2 t^{-3} + C_\ST^2t^{-6/5}, \Gamma_{\sh,\mathrm{max}}^{2})
\end{align}
where $\beta_\sh$ is the shock speed divided by the speed of light $c$, which is measured in the laboratory frame, $\Gamma_\sh = 1/\sqrt{1 - \beta_\sh^2}$ is the shock Lorentz factor, and $t$ denotes the elapsed laboratory time since the explosion. The coefficients $C_\BM$ and $C_\ST$ are given by
\begin{align}
\label{eq.C_BM}
C_\BM &= \sqrt{\frac{17E}{8\pi n_0\mpr c^5}},\\
\label{eq.C_ST}
C_\ST &= \frac{2}{5}\cdot 1.15 \left(\frac{E}{n_0 \mpr c^5}\right)^{1/5},
\end{align}
where $n_0$ and $\mpr$ stand for the ambient number density and the proton mass, respectively.\footnote{We use $E$ to define $C_\ST$ in Equation~\eqref{eq.C_ST} instead of $E\theta_\mathrm{j}^2/2$, where $\theta_\mathrm{j}$ is the jet truncation angle, which was used in equation~(A10) in \citet{Eerten10}.
We do not follow \citet{Eerten10}, since $E\theta_\mathrm{j}^2/2$ does not have a clear physical meaning for a structured jet, although it means the jet energy contained in the approaching and counter jets for a uniform jet. The use of $E$ in Equation~\eqref{eq.C_ST} is consistent with our assumption that each jet segment radially expands as if it were a portion of the isotropic blast wave with the same isotropic equivalent energy. Our conclusions do not change even if $E$ in Equation~\eqref{eq.C_ST} is replaced by $E\theta_\mathrm{j}^2/2$.}
The numerical factor $1.15$ in Equation~(\ref{eq.C_ST}) originates from the energy conservation. 
$\Gamma_{\sh,\mathrm{max}}=100$ is the shock Lorentz factor before the shock deceleration phase.
The radius of each shock segment at $t$ is given by integrating the shock speed:
\begin{align}
\label{eq.R}
R = \int _0^t c\beta_\sh\diff t.
\end{align}

Equations~\eqref{eq.BM-STsh}-\eqref{eq.R} determine the shock dynamics described by the shock Lorentz factor $\Gamma_\sh(t,\theta)$ and shock radius $R(t,\theta)$, once a jet structure $E(\theta)$ and ambient number density $n_0$ are given. 

\subsubsection{Particle acceleration \& shock downstream quantities}\label{sec.method.p}

\begin{figure}
	\includegraphics[width = \columnwidth]{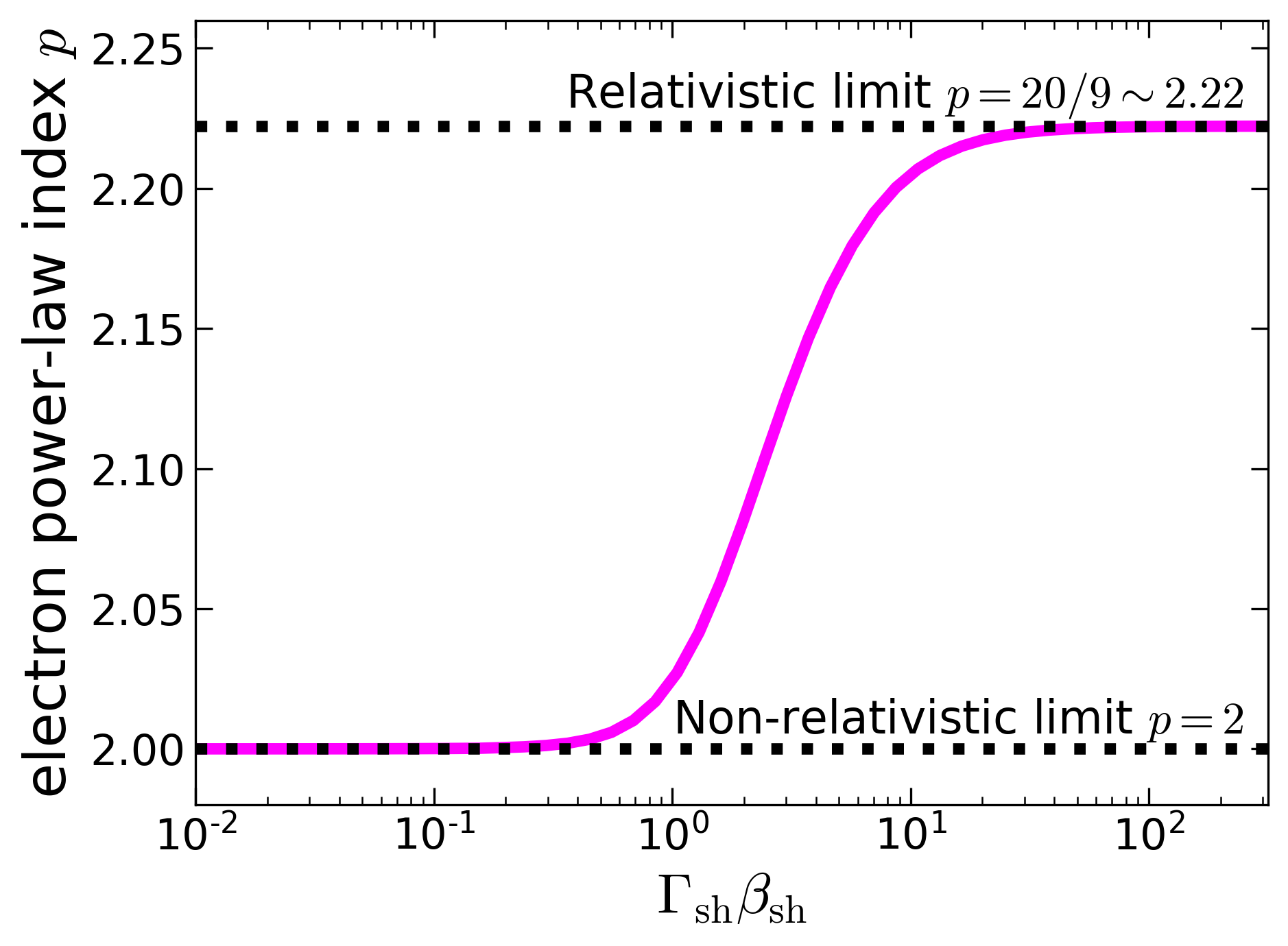}
	\caption{Electron power-law index $p$ as a functioregion that emits the photons n of $\Gamma_\sh \beta_\sh$, which is given by Equation~(\ref{eq.p}). The line is fully consistent with the green solid line in figure~1 in \citet{KW05}. The upper dotted line shows the relativistic limit $p=20/9\sim 2.22$ ($\Gamma_\sh \gg 1$) while the lower one shows the non-relativistic limit $p=2$ ($\Gamma_\sh\rightarrow 1$).}
	\label{fig.p}
\end{figure}

The physical quantities in the downstream flow just below the shock wave are determined from the Rankine-Hugoniot conditions for $\Gamma_\sh$ given by Equation~(\ref{eq.BM-STsh}). The shock jump conditions for a cold, stationary upstream flow are summarized as \citep[e.g.,][]{BM}
\begin{align}
    \label{eq.RH1}
    \Gamma_\sh^2 &= \frac{(\Gamma + 1)[1 + \hat{\gamma}'(\Gamma - 1)]^2}{2+\hat{\gamma}'(2-\hat{\gamma}')(\Gamma - 1)},\\
    \label{eq.RH2}
    \frac{n'}{n_0} &= \frac{\hat{\gamma}'\Gamma + 1}{\hat{\gamma}' - 1},\\
    \label{eq.RH3}
    \frac{e_{\mathrm{i}}'}{n'\mpr c^2} &= \Gamma - 1,
\end{align}
where $\Gamma=1/\sqrt{1-\beta^2}$ is the Lorentz factor of the shocked fluid with $c\beta$ being the speed of the shocked flow, $n'$ the downstream number density, $\hat{\gamma}'$ the ratio of the specific heats of the shocked fluid, and $e_\mathrm{i}'$ the downstream internal energy density. 

We incorporate the \Juttner-Synge equation of state (EoS) \citep{Juttner,Synge,RelaBoltzmann} with Equations~(\ref{eq.RH1})-(\ref{eq.RH3}), which follows from the isotropic equilibrium state of a non-degenerate gas by taking into account the relativistic effects:
\begin{align}
    \label{eq.JS1}
    P' &= n'k_\mathrm{B}\mathcal{T}' = (\hat{\gamma}' - 1) e_\mathrm{i}',\\
    \label{eq.JS2}
    \frac{h'}{c^2} &= 1 + \frac{e_\mathrm{i}' + P'}{n' \mpr c^2} = \frac{\displaystyle K_3\left(\zeta \right)}{\displaystyle K_2\left(\zeta \right)},
\end{align}
where $P'$ is the gas pressure, $k_\mathrm{B}$ the Boltzmann constant, $h'$ the specific enthalpy including the rest mass contribution, and $K_j$ the modified Bessel function of the second kind.
A dimensionless quantity $\zeta$ is defined by
\begin{align}
    \zeta = \frac{\mpr c^2}{k_\mathrm{B}\mathcal{T}'},
\end{align}
where $\mathcal{T}'$ is the gas temperature.
We note that the ratio of the specific heats $\hat{\gamma}'$ is given by Equations~\eqref{eq.JS1} and \eqref{eq.JS2} as 
\begin{align}
    \label{eq.zeta_hatgamma}
    \hat{\gamma}' = 1+\frac{P'}{e_{\mathrm{i}}'} =  1 + \frac{4 K_2(\zeta)}{\zeta[3K_3(\zeta) + K_1(\zeta) - 4K_2(\zeta)]},
\end{align}
where $\hat{\gamma}' \rightarrow 4/3$ in the relativistic limit ($\zeta \rightarrow 0$) while $\hat{\gamma}' \rightarrow 5/3$ in the non-relativistic limit ($\zeta \rightarrow \infty$).

In practice, we obtain the downstream quantities $\Gamma$, $n'$, and $e_\mathrm{i}'$ as follows. Noting that the downstream Lorentz factor $\Gamma$ is given by Equations~\eqref{eq.RH3}-\eqref{eq.JS2} as
\begin{align}
    \label{eq.zeta_Gamma}
    \Gamma &= \frac{K_3(\zeta)}{K_2(\zeta)} - \frac{1}{\zeta},
\end{align}
we substitute Equations~(\ref{eq.zeta_hatgamma}) and (\ref{eq.zeta_Gamma}) to Equation~(\ref{eq.RH1}) and numerically find the solution $\zeta$ that satisfies the equation for a given $\Gamma_\sh$. The solution $\zeta$ gives $\hat{\gamma}'$ and $\Gamma$ through Equations~(\ref{eq.zeta_hatgamma}) and (\ref{eq.zeta_Gamma}), respectively.
Then, the downstream number density $n'$ is determined from Equation~(\ref{eq.RH2}) for a given $n_0$. Lastly, Equation~(\ref{eq.RH3}) gives the downstream internal energy density $e_\mathrm{i}'$. The other downstream quantities are given by these quantities.
We also note that Equations~(\ref{eq.RH1})-(\ref{eq.zeta_Gamma}) recover the well-known jump conditions both in the relativistic ($\Gamma_\sh \gg 1$ or $\zeta \rightarrow 0$) and non-relativistic ($\Gamma_\sh \rightarrow 1$ or $\zeta \rightarrow \infty$)  limits and smoothly connect them as noted in Appendix~\ref{app.downstream}.

The non-thermal electrons are assumed to be produced via diffusive shock acceleration at the forward shock and possess an isotropic energy spectrum given by a single power law with an index $p$. 
When the synchrotron cooling effect can be neglected as in the slow cooling phase, the electron energy distribution function $N_\mathrm{e,s}'(E_\mathrm{e}')$ is given by
\begin{align}
    \label{eq.e-dist}
    N_\mathrm{e,s}'(E_\mathrm{e}') \diff E_\mathrm{e}' &= 
    \begin{cases}
    C_\mathrm{s} \gamma_\mathrm{e}^{\prime-p} \diff \gamma_\mathrm{e}' & (\gamma_\mathrm{e}' \ge \gamma_\mathrm{m}')\displaystyle \\
    0 & (\gamma_\mathrm{e}' < \gamma_\mathrm{m}')
    \end{cases},\\
    \label{eq.C}
    C_\mathrm{s} &= (p-1)n_\mathrm{R}^\prime \gamma_\mathrm{m}^{\prime p-1},
\end{align}
where $E_\mathrm{e}'$ is the energy of an electron.
The minimal Lorentz factor of the non-thermal electrons $\gamma_\mathrm{m}'$ and the number density of the relativistic electrons, which are responsible for synchrotron radiation, $n_\mathrm{R}^\prime$ are given by \citep{Sari98,SG13}
\begin{align}
    \label{eq.gmin}
    \gamma_\mathrm{m}' &= \max \left(1,\  \frac{p-2}{p-1}\frac{\mpr}{m_\mathrm{e}}\ee(\Gamma-1)\right), \\
    \label{eq.nR}
    n_\mathrm{R}^\prime &= n^\prime \min\left(1,\
    \frac{p-2}{p-1}\frac{\mpr}{m_\mathrm{e}}\ee(\Gamma-1)
    \right),
\end{align}
with $\ee$ being the energy conversion fraction from the shocked matter to the accelerated electrons, which are valid for $p$ given by Equation~\eqref{eq.p} below.\footnote{We assume that $\ee$ does not evolve and remains constant, since $\ee$ has a narrow distribution centered around $\ee \sim 0.14$ for observational data collected at different times, which implies that $\ee$ does not strongly depend on the shock Lorentz factor \citep{ee17}.}
Equations~\eqref{eq.gmin} and \eqref{eq.nR} come from the assumption that the spectrum of accelerated electrons follows a power-law distribution in momentum with the index $p$ and
take into account the effect that the number of the relativistic electrons becomes subdominant compared to the non-relativistic electrons, while the energy of the relativistic electrons remains dominant, as the shock is decelerated to a non-relativistic speeds.

We assume that the power-law index $p$ depends on the shock speed.
This study employs the model by \citet{KW05} as an example, who derived the following equation by considering the diffusive shock acceleration across a relativistic shock wave with an isotropic diffusion:
\begin{align}
    \label{eq.p}
    p = \frac{3\betau - 2\betau \betad^2 + \betad^3}{\betau - \betad} -2,
\end{align}
where $\betau$ and $\betad$ are respectively the upstream and downstream fluid speeds measured in the shock rest frame in unit of $c$.\footnote{We note that Equation~\eqref{eq.p} does not explain a wide range of $p$ measured in GRB afterglows, while it could explain the observation in GRB~170817A as shown below (see also Section~\ref{sec.disc.rate} for discussion).}
Through the Lorentz transformation, we find
\begin{align}
    \betau &= \beta_\sh,\\
    \betad &= \frac{\beta_\sh - \beta}{1 - \beta \beta_\sh}.
\end{align}
We note that the value of $p$ is different among the jet segments, since $\Gamma_\sh(t, \theta)$ evolves in different ways for each jet segment, depending on the jet structure $E(\theta)$.
Equation~(\ref{eq.p}) yields $p=2$ in the non-relativistic limit $(\Gamma_\sh \rightarrow 1)$ and $p=20/9 \sim 2.22$ in the relativistic limit $(\Gamma_\sh \gg 1)$ with the \Juttner-Synge EoS \citep{KW05}. Figure~\ref{fig.p} depicts the electron power-law index $p$ as a function of $\Gamma_\sh \beta_\sh$.

When the synchrotron cooling is effective as in the fast cooling phase, the electron energy distribution function is given by
\begin{align}
    \label{eq.e-dist2}
    N_\mathrm{e,f}'(E_\mathrm{e}') \diff E_\mathrm{e}' &= 
    \begin{cases}
    C_\mathrm{f} \gamma_\mathrm{e}^{\prime-2} \diff \gamma_\mathrm{e}' & (\gamma_\mathrm{e}' \ge \gamma_\mathrm{c}')\displaystyle \\
    0 & (\gamma_\mathrm{e}' < \gamma_\mathrm{c}')
    \end{cases},\\
    \label{eq.C2}
    C_\mathrm{f} &= n_\mathrm{R}^\prime \gamma_\mathrm{c}^{\prime},
\end{align}
where
\begin{align}
    \gamma_\mathrm{c}^{\prime} = \frac{3\me c \Gamma}{4\sigma _\mathrm{T} \eB e_\mathrm{i}' t},
\end{align}
is the characteristic Lorentz factor of the electrons for synchrotron cooling, with $\sigma_\mathrm{T}$ being the cross section of Thomson scattering.

The strength of the downstream magnetic field is phenomenologically given by
\begin{align}
    \label{eq.B}
    B' = \sqrt{8\pi \eB e'_\mathrm{i}},
\end{align}
where $\eB$ is a parameter that gives the energy conversion fraction from the shocked matter to the magnetic field. The downstream magnetic field is assumed to be well tangled, which results in isotropic synchrotron emission in the fluid rest frame as assumed below.

\subsubsection{Observed synchrotron flux}\label{sec.method.obsFlux}
The observed afterglow is calculated by integrating the synchrotron radiation emitted from each position on the jet surface. 
The local synchrotron emissivity at the fluid rest frame is evaluated based on the standard model of GRB afterglows \citep{Sari98}. The synchrotron radiation is assumed to be isotropic in the fluid rest frame as a result of a well-tangled magnetic field in the shocked flow. The synchrotron emissivity at the fluid rest frame is well approximated by a broken power law bent at the synchrotron characteristic frequency $\numd$ and cooling frequency $\nucd$. In the case of slow cooling ($\numd < \nucd$), the energy radiated by synchrotron emission per unit volume per unit time per unit frequency, $\emisd$, is given by
\begin{align}
\label{eq.slowcooling}
\emisd = \epsilon '_{\nu',\mathrm{p}}\times \left\{
\begin{array}{ll}
\displaystyle \left(\frac{\nu'}{\numd}\right)^{1/3} & (\nu' < \numd)\\
\displaystyle \left(\frac{\nu'}{\numd}\right)^{-(p-1)/2} & (\numd \le \nu' < \nucd) \\
\displaystyle \left(\frac{\nucd}{\numd}\right)^{-(p-1)/2}\left(\frac{\nu'}{\nucd}\right)^{-p/2} & (\nucd \le \nu')
\end{array} \right.. 
\end{align}
In the case of fast cooling ($\numd \ge \nucd$), on the other hand, $\emisd$ is given by
\begin{align}
\label{eq.fastcooling}
\emisd = \epsilon '_{\nu',\mathrm{p}}\times \left\{
\begin{array}{ll}
\displaystyle \left(\frac{\nu'}{\nucd}\right)^{1/3} & (\nu' < \nucd)\\
\displaystyle \left(\frac{\nu'}{\nucd}\right)^{-1/2} & (\nucd \le \nu' < \numd) \\
\displaystyle \left(\frac{\numd}{\nucd}\right)^{-1/2}\left(\frac{\nu'}{\numd}\right)^{-p/2} & (\numd \le \nu')
\end{array} \right.. 
\end{align}
The peak emissivity in Equations~\eqref{eq.slowcooling} and \eqref{eq.fastcooling}
is given by \citep{Granot99,Eerten10}
\begin{align}
\label{eq.emisdpeak}
\epsilon '_{\nu',\mathrm{p}} 
= 0.88 \cdot \frac{256}{27}\frac{p-1}{3p-1}\frac{\qe^3}{\me c^2} n_\mathrm{R}^\prime B',
\end{align}
where $\qe$ is the elementary charge. The numerical factor $0.88$ is for better fitting the broken power-law spectrum to the exact spectrum.
The two break frequencies are given by \citep{Granot99,Eerten10}
\begin{align}
\label{eq.numd}
\numd &=  \frac{3}{16}  \frac{\gamma _\mathrm{m}^{\prime 2}\qe B'}{\me c}, \\
\label{eq.nucd}
\nucd &=  \frac{3}{16}  \frac{\gamma _\mathrm{c}^{\prime 2}\qe B'}{\me c} = \frac{3}{16} \left[\frac{3\me c \Gamma}{4\sigma _\mathrm{T} \eB e_\mathrm{i}' t }\right]^2 \frac{\qe B'}{\me c}.
\end{align}
The coefficient $3/16$ results from eq.~(6.17c) in \citet{RL} by assuming the isotropic distribution of the electron pitch angle and taking the pitch-angle average.

The observed synchrotron flux is given by
\begin{align}
    \label{eq.Fnu}
    F_\nu (T)&= \frac{1}{4\pi D^2} \int _0^{\theta_\mathrm{j}} \diff \theta \sin \theta 
    \int _0^{2\pi}\diff \varphi \nonumber \\ 
    & \quad \times \left. \frac{|\mu - \beta_\sh|}{1-\mu\beta_\sh}\frac{R^2}{\Gamma^3(1-\beta \mu)^3}\frac{\emisd}{\alpha'_{\nu'}}(1-e^{-\tau_\nu})\right|_{t=T+\mu R/c},
\end{align}
under the thin-shell approximation (See Appendix~\ref{app.eq} for the derivation of Equation~\eqref{eq.Fnu}). $D$ is the luminosity distance to the source, $\theta_\mathrm{j}$ is the jet truncation angle specified by a given jet structure, $\alpha'_{\nu'}$ is the absorption coefficient of synchrotron self-absorption given by Equation~\eqref{eq.alpha_fin}, and $\tau_\nu$ is the optical depth given by Equation~\eqref{eq.opacity}.
$\mu$ is the cosine of the angle between the radial direction and the line of sight. For an observer at $(\theta,\varphi)=(\view,0)$, where $\view$ is a given viewing angle and $\varphi$ is the azimuthal angle around the jet axis, $\mu$ is given by
\begin{align}
    \label{eq.mu}
    \mu = \sin \theta \sin \view \cos \varphi + \cos \theta \cos \view.
\end{align}
The factor $1/[\Gamma^3(1-\beta \mu)^3]$ comes from the Lorentz transformation of $\epsilon_\nu/\alpha_\nu$. The integrand is evaluated at the laboratory time when the emitted photons reach the observer at an observer time $T$:
\begin{align}
    \label{eq.t}
    t = T+\frac{\mu R}{c},
\end{align}
where we chose $T = 0$ as the arrival time of a photon emitted at the origin at $t = 0$.
The observed frequency $\nu$ and the rest frame frequency $\nu'$ are related via the Lorentz transformation as
\begin{equation}
\label{eq.nud}
\nu' = \Gamma(1-\beta \mu)\nu .
\end{equation}

The observed afterglow flux for a given frequency $\nu$ is calculated by Equation~(\ref{eq.Fnu}) after specifying the parameters $\{ n_0, \eB, \ee, \view, D\}$ and a jet structure $E(\theta)$.

\subsection{Model of structured jets}\label{sec.method.jet}

\begin{figure}
	\includegraphics[width = \columnwidth]{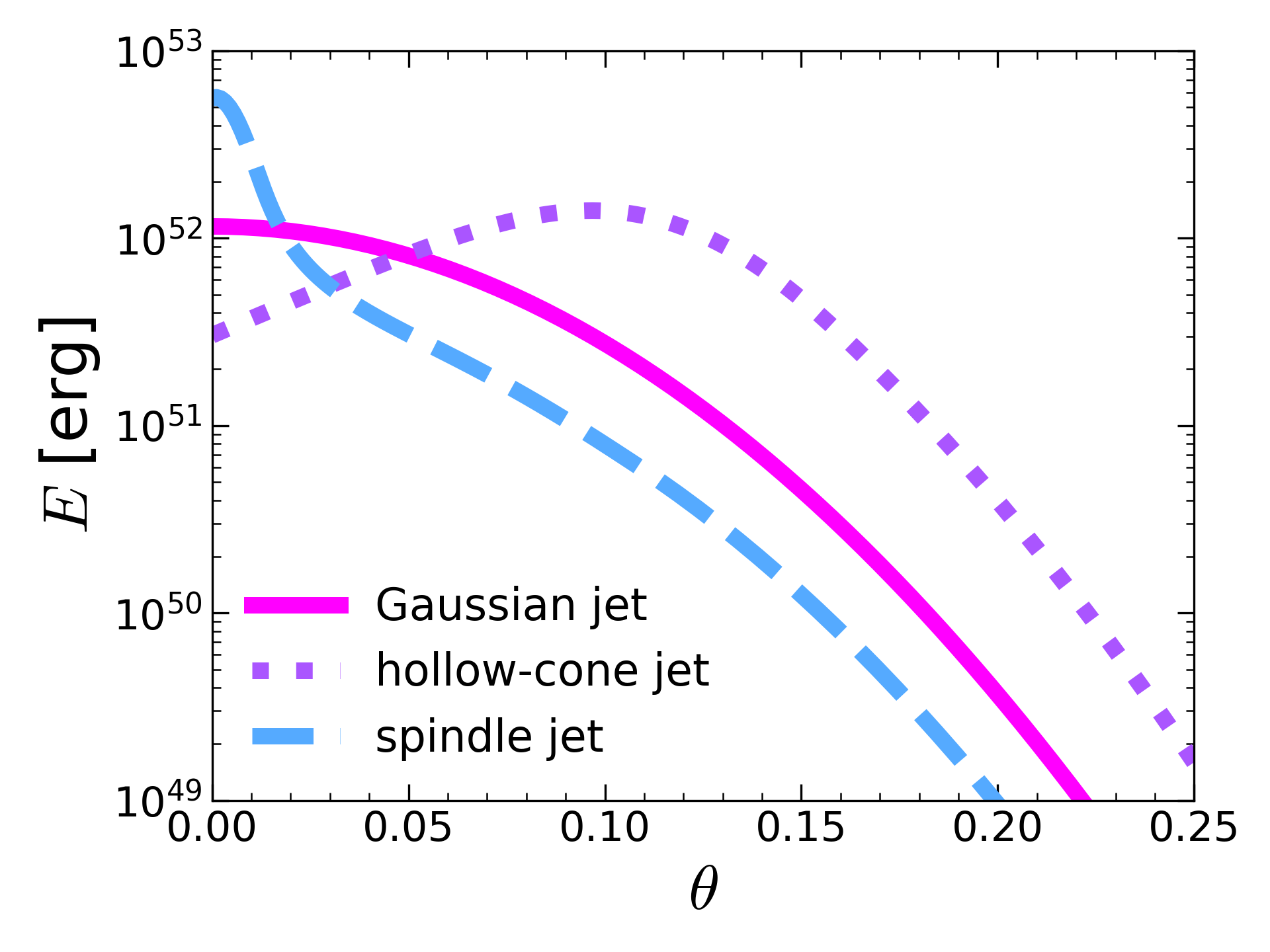}
	\caption{Three kinds of the jet structures studied in this paper, which are referred from figure~1 in \citetalias{TI21}:
	a hollow-cone jet (dotted line), a Gaussian jet (solid line), and a spindle jet (dashed line).
	The Gaussian jet is given by $E(\theta) = E_\mathrm{G} \exp [-\theta^2/(2\theta_\mathrm{G}^2)]$
	with $E_\mathrm{G} = 1.16\times 10^{52}$~erg and $\theta_\mathrm{G}=0.059$.
	The jet is truncated at $\theta = \theta_\mathrm{j} = 0.3$.
	They are consistent with the afterglow of GRB~170817A as shown in \citetalias{TI21}.}
	\label{fig.jet}
\end{figure}

Observed afterglow fluxes are calculated for a given jet structure $E(\theta)$ as explained in Section~\ref{sec.method.afterglow}.
In this study, we use the jet structures presented in Figure~\ref{fig.jet},
which shows three examples of the structured jets that generate afterglows consistent with those of GRB~170817A within observational errors under the assumption of a constant value of the electron power-law index $p=2.17$ and $n_0=10^{-3}$~cm$^{-3}$: hollow-cone, Gaussian, and spindle jets, which are referred from figure~1 in \citetalias{TI21}.\footnote{We truncate the jet edge at $\theta = \theta_\mathrm{j}=0.3$ in this study while $\theta _\mathrm{j} = 0.61$ in the original paper \citepalias{TI21}, in order to save the numerical costs for calculating shock dynamics and afterglow emissions at the jet edge. This is justified because the emission from the jet edge $\theta > 0.3$ does not affect the observed afterglow fluxes at the epochs focused on in this paper.}
This study employs these non-uniform GRB jets, since at least GRB~170817A rejects a uniform top-hat jet as shown by afterglow observations \citep{Mooley18,Troja18a}.
There are also other candidates of the jet structure of GRB~170817A, which include
Gaussian jets \citep{Lyman18,Resmi18,Troja19,LK18,rapiddeclineHST,GG20,Ryan,Troja20}, power-law jets \citep{LK18,DAvanzo18,Ghirlanda19,Beniamini20,GG20,Ryan}, and hollow-cone jets with a power-law edge \citepalias{TI20}.
However, we find below that our main results do not much depend on the jet structure qualitatively nor quantitatively. Hence, it would be enough that the study focuses on the jet structures in Figure~\ref{fig.jet}.

\section{Results}\label{sec.results}

\begin{figure*}
	\includegraphics[width=\textwidth]{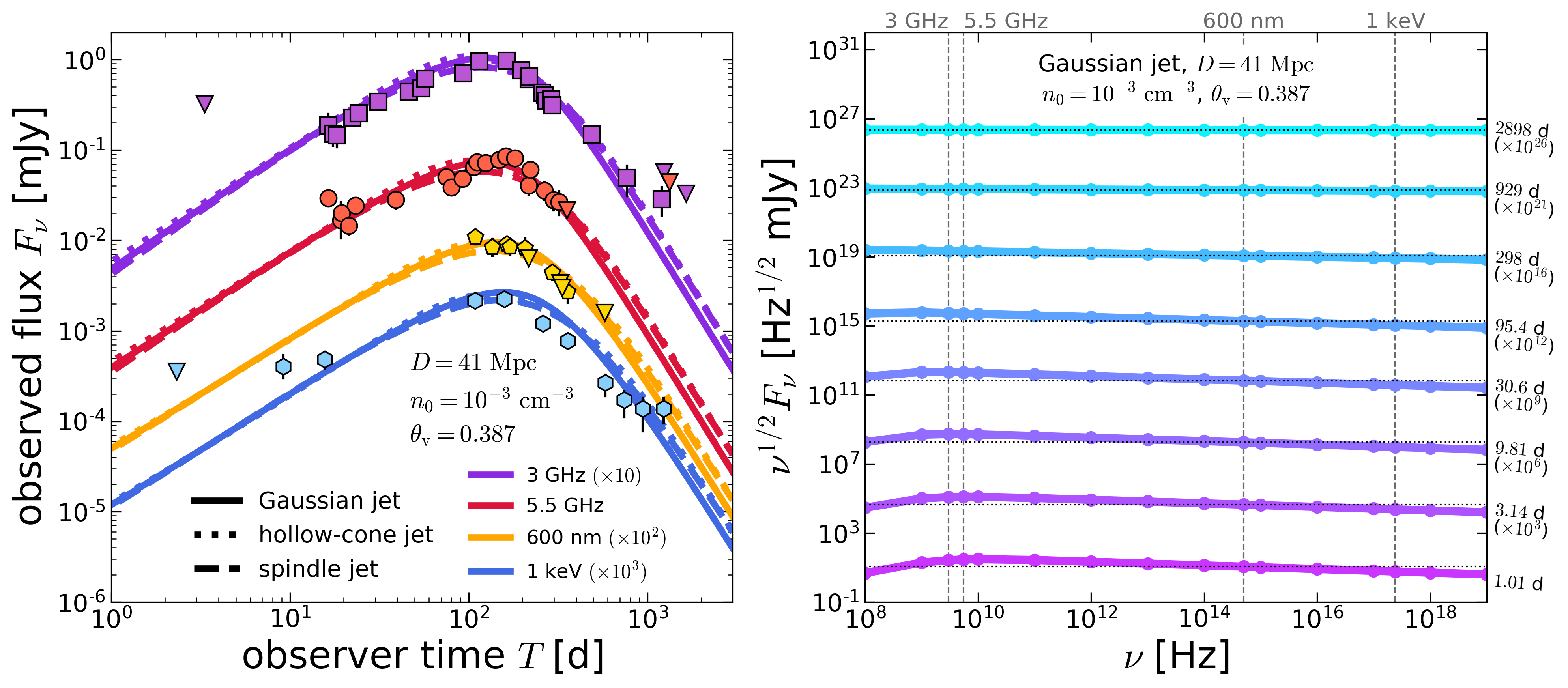}
	\caption{
	{\bf Left:} Afterglow light curves of the structured jets with the electron power-law index $p$ that depends on the shock Lorentz factor as given by Equation~(\ref{eq.p}). The solid, dotted, and dashed lines show the light curves of the Gaussian, hollow-cone, and spindle jets, respectively.
	Each colour corresponds to the radio (3~GHz, purple; 5.5~GHz, red), optical (600~nm, orange), and X-ray (1~keV, blue).
    Also plotted are the observed afterglow fluxes (points) and upper limits (lower triangles) of GRB~170817A. 
    The flux values for 3~GHz are quoted from \citet{Panchromatic,Balas,Troja21,Balas22}; 
    Those for 600~nm are taken from \citet{rapiddeclineHST} and \citet{Kilpatrick} while those for 1~keV are cited from \citet{Troja21}.
    The data points for 5.5~GHz are from Figure~4 in \citet{Troja19}, which uses the data in \citet{Hallinan17,Lyman18,Troja18a,Margutti18,Mooley18,Alexander18,Piro19}, while the upper limit at $T=1330$~d is from \citet{Troja21}.
    {\bf Right:} Afterglow spectra of the Gaussian jet, where we multiply $\nu^{1/2}$ by $F_\nu$. The spectra are arranged in chronological order from bottom to top by multiplying a factor to each spectrum as designated with the observer time in the right side of the figure. Each dotted line is horizontal, on which $F_\nu \propto \nu^{-1/2}$. 
	}
	\label{fig.Gaussian_fid}
\end{figure*}

\begin{figure*}
	\includegraphics[width=\textwidth]{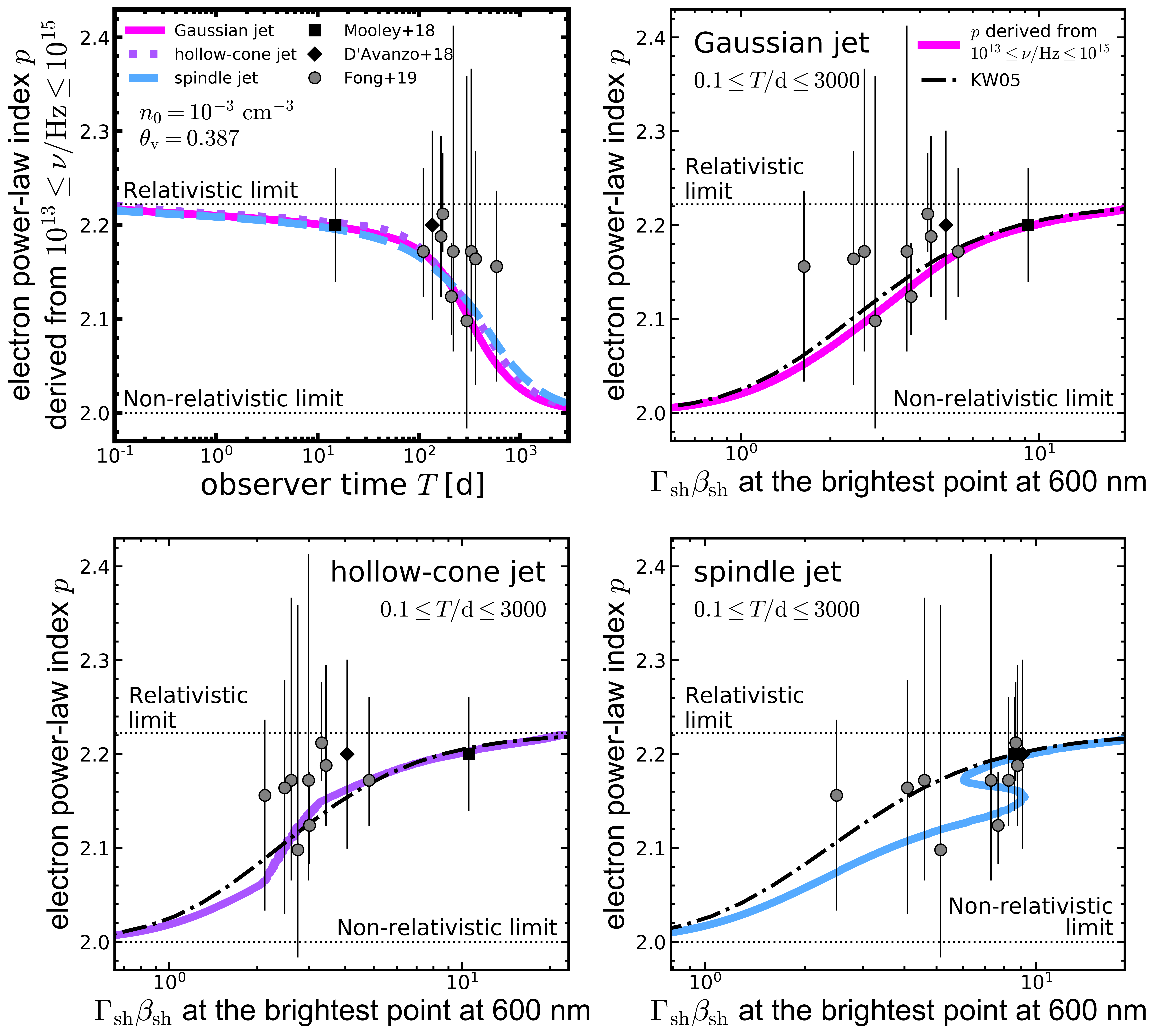}
	\caption{
	{\bf Upper left:}
	Evolution of the electron power-law index $p$ obtained from the calculated afterglow spectra in $10^{13}\le \nu/\mathrm{Hz}\le 10^{15}$ for various jet structures.
	The upper and lower dotted lines show the relativistic limit $(p=2.22)$ and the non-relativistic limit $(p=2)$ of the diffusive shock acceleration model given by Equation~(\ref{eq.p}).
	Also plotted are the observed electron power-law index in GRB~170817A that are derived from multi-band spectra.
	The square-shaped plot is referred from \citet{Mooley18}, which is derived from radio and X-ray fluxes.
	The diamond-shaped plot is referred from \citet{DAvanzo18}, which is derived from optical and X-ray fluxes.
	The round-shaped plots are referred from table~2 in \citet{Fong19}, which uses the radio, optical, and X-ray data in \citet{Alexander18,Margutti18,Mooley18,rapiddecline,Dobie18,Hajela19,Piro19,Troja19}. Uncertainties correspond to $1\sigma$.
	{\bf The others:}
	Evolution of the electron power-law index $p$ obtained from the calculated spectrum as a function of $\Gamma_\sh \beta_\sh$ measured at the point with the largest $\diff F_\nu/\diff \Omega$ ($\nu = 500$~THz) on the jet surface for the Gaussian {\bf(Upper right)}, hollow-cone {\bf(Lower left)}, and spindle jets {\bf(Lower right)}. The presented range of $\Gamma_\sh \beta_\sh$ corresponds to $0.1 \le T/\mathrm{d}\le 3000$ for each jet model. The dot-dashed line in each panel indicates the relation between the electron power-law index given by Equation~\eqref{eq.p} and $\Gamma_\sh \beta_\sh$, which is the same as in Figure~\ref{fig.p}. The observational data points are the same as in the upper-left panel but we converted the observer time $T$ to the corresponding $\Gamma_\sh \beta_\sh$ for each jet model.
	}
	\label{fig.structure_independence}
\end{figure*}

We investigate the time evolution and detectability of the electron power-law index $p$ inferred from GRB afterglow spectra, by calculating afterglow fluxes and spectra for three non-uniform structured jets. 
First, we show that the employed particle acceleration model of \citet{KW05} is consistent with the observed spectral evolution of the afterglow of GRB~170817A within observational errors in Section~\ref{sec.result_structure}.
We also find that the time evolution of the electron power-law index $p$ obtained from the calculated spectrum does not depend on the jet structure.
Next, we show that there would be a chance to probe the particle acceleration model 
by future observations of off-axis GRB afterglows
in Section~\ref{sec.result_n0}.

\subsection{Results for a canonical parameter set for GRB~170817A}\label{sec.result_structure}
Figures~\ref{fig.Gaussian_fid} and \ref{fig.structure_independence} demonstrate that 
the afterglow of GRB~170817A is still consistently explained even if we take into account the evolution of the electron power-law index $p$ introduced by Equation~\eqref{eq.p}, where
Figure~\ref{fig.Gaussian_fid} shows the calculated afterglow light curves and spectral evolution
and Figure~\ref{fig.structure_independence} shows the evolution of the electron power-law index $p$ obtained from the calculated afterglow spectrum.
We generated the afterglow by using Equation~\eqref{eq.Fnu} and the structured jet models in Figure~\ref{fig.jet} with the fiducial parameter values for GRB~170817A adopted in \citetalias{TI21}:
$n_0=10^{-3}$~cm$^{-3}$, 
$\view=0.387\sim22.2^\circ$,
$\ee=0.1$, 
and 
$D=41$~Mpc, while $\eB$ is tuned for each jet structure as
$\eB=2.44\times10^{-4}$ for the Gaussian jet,
$\eB=4.12\times10^{-5}$ for the hollow-cone jet,
and  
$\eB=6.60\times10^{-4}$ for the spindle jet.

The left panel of Figure~\ref{fig.Gaussian_fid} confirms that the afterglow light curves for four bands (3~GHz, 5.5~GHz, 600~nm, and 1~keV) are consistent with the observed ones within observational errors, irrespective of the jet structure.
This would be an expected result, since the change of $p$ introduced by Equation~\eqref{eq.p} is limited in $2 \le p \le 2.22$, which is not much different from the value adopted in our previous paper ($p=2.17$, \citetalias{TI21}).

The right panel in Figure~\ref{fig.Gaussian_fid} displays the corresponding afterglow spectra of the Gaussian jet. The spectra for the hollow-cone and spindle jets are almost the same as those for the Gaussian jet and, hence, are omitted. 
Each dotted line is horizontal and indicates $F_\nu \propto \nu^{-1/2}$ or, in other words, $p=2$ for $\nu_\mathrm{m} \le \nu < \nu_\mathrm{c}$ (cf.~Equation~\ref{eq.slowcooling}). 
As seen in the panel, the value of $p$ gradually decreases and approaches $p=2$ as time passes.
Note that we implicitly assumed here that an observed spectrum has the same spectral slope as that in the fluid rest frame (Equation~\ref{eq.slowcooling}), whereas the observed spectrum is produced by integrating the local emissions on the jet surface. 
This assumption is justified for off-axis GRB afterglows, because the observed afterglow is dominated by a small region around the brightest point on the jet surface, which gradually moves from the jet edge region to the inner region, due to the relativistic beaming effect \citepalias{TI20}. 
Hence, the observed spectrum has almost the same spectral slope as that of the spectrum of the emission from the small luminous region. That is, the observed logarithmic spectral slope in $\nu_\mathrm{m} \le \nu < \nu_\mathrm{c}$ is given by $-(p-1)/2$, where $p$ is the electron power-law index given by Equation~\eqref{eq.p} with a typical shock Lorentz factor for the small region.
Here, we can put $\nu_\mathrm{m} = \numd/[\Gamma (1-\beta \mu)]$ and $\nu_\mathrm{c} = \nucd/[\Gamma (1-\beta \mu)]$, where $\numd$, $\nucd$, $\mu$, $\Gamma$ and $\beta$ are measured at the most luminous point, by the same reason.

The upper-left panel of Figure~\ref{fig.structure_independence} shows that the time evolution of the electron power-law index $p$ obtained from the calculated spectrum is consistent with observations, irrespective of the jet structure.
Here, we derived the electron power-law index $p$ from the spectral slope in $10^{13} \le \nu/\mathrm{Hz} \le 10^{15}$, which was so conservatively chosen to avoid the effects of the spectral breaks at $\nu_\mathrm{m}$ and $\nu_\mathrm{c}$.
The evolutionary paths for the three jet structures are similar to each other, which reflects the similarity of the light curves in Figure~\ref{fig.Gaussian_fid}.
The power-law index obtained from the calculated spectrum rapidly transits from the relativistic limit $p=2.22$, to the non-relativistic limit, $p=2$, roughly at the peak time of the afterglow fluxes, $T \sim 100$~d.
The transition takes place at around the peak time because the afterglow peak is dominated by 
the most energetic region of the jet, where $E(\theta)$ is the maximum, \citepalias[see fig.~5 in][]{TI21} and the region becomes observable after the shock slows down to a trans-relativistic speed.
The photons emitted from the most energetic part can reach the off-axis observer only after the shock is decelerated to 
$\Gamma_\sh \sim \sqrt{2} \Gamma \sim \sqrt{2}(\view - \theta_\mathrm{E})^{-1} \sim 4\mathchar`-5$,
where we used Equation~(\ref{eq.strong_rela_Gamma}) and 
$\theta_\mathrm{E}$ denotes the angle of the most energetic part ($\theta_\mathrm{E}=0$ for the Gaussian and spindle jets while $\theta_\mathrm{E}\sim 0.1$ for the hollow-cone jet; see Figure~\ref{fig.jet}). Around the shock Lorentz factor $\Gamma_\sh \sim 4\mathchar`-5$, the electron power-law index introduced by Equation~(\ref{eq.p}) rapidly transits between $p=2.22$ and $p=2$ as shown in Figure~\ref{fig.p}.

The upper-right and lower-left panels in Figure~\ref{fig.structure_independence} demonstrate 
that the spectral slopes of the afterglows are basically determined by the electron power-law index introduced by Equation~\eqref{eq.p} at the brightest point of the shock surface as mentioned above for the Gaussian and hollow-cone jets.
In these panels, we plotted the electron power-law index obtained from the calculated spectrum as a function of $\Gamma_\sh \beta_\sh$ measured at the point on the jet surface that is the brightest at the wavelength of $600$~nm in term of the flux per unit solid angle $\diff F_\nu/\diff \Omega$.\footnote{We used $\Gamma_\sh \beta_\sh$ at the brightest point instead of time, since $\Gamma_\sh \beta_\sh$ at the brightest point monotonically decreases with time as the emission centroid moves from the jet edge part to the inner region in the case of the Gaussian and hollow-cone jets. The monotonicity does not hold in the case of the spindle jet as explained in the next paragraph.}
The relation between the electron power-law index $p$ obtained from the calculated spectrum and $\Gamma_\sh \beta_\sh$ at the brightest point (a coloured solid line) agrees well with the relation given by Equation~(\ref{eq.p}) (a black dot-dashed one), which results from the fact that the brightest point dominates the observed flux and determines the observed spectral shape.
The flux contribution from the region around the brightest point causes a slight difference between the electron power-law index obtained from the calculated spectrum and that given by Equation~\eqref{eq.p}: The former becomes larger (smaller) than the latter if the shock Lorentz factor of the surrounding contributing region is larger (smaller).

The lower-right panel in Figure~\ref{fig.structure_independence} shows the same result but for the spindle jet.
It is the same as for the Gaussian and hollow-cone jets that a small luminous region around the brightest point dominates the observed flux, but the power-law index obtained from the calculated spectrum does not represent the electron power-law index introduced by Equation~\eqref{eq.p} at the brightest point.
The relation for the electron power-law index obtained from the calculated spectrum is given by an S-shaped curve (the solid curve) and is different from the relation given by Equation~\eqref{eq.p} (the dot-dashed curve).
This is because the $\Gamma_\sh \beta_\sh$ measured at the brightest point does not monotonically decrease with time but increases once the emission centroid approaches the central energetic region around the jet axis ($\theta \lesssim 0.05$, see Figure~\ref{fig.jet}), while the electron power-law index obtained from the calculated spectrum decreases with time as seen in the upper-left panel of Figure~\ref{fig.structure_independence}.
In the previous case of the Gaussian and hollow-cone jets, the brightest point can enter the inner region only after the Lorentz factor at the inner region drops below the value outside. However, in the case of the spindle jet, the brightest point enters the central energetic region with a larger Lorentz factor, since the enhancement of the observed flux by the larger jet energy overcomes the suppression by the stronger relativistic debeaming. 
After the brightest point reaches near the jet axis where the energy distribution is rather flat ($\theta \sim 0.01$), $\Gamma_\sh \beta_\sh$ at the brightest point decreases with time again. 
On the other hand, the electron power-law index obtained from the calculated spectrum monotonically declines with time, since the observed flux is dominated by the contribution from the less energetic region outside the brightest point with a low $\Gamma_\sh \beta_\sh$, where its typical shock 4-velocity continues to decrease with time.

As a short summary, the time evolution of the electron power-law index $p$ obtained from afterglow spectra is still consistent with the observations of GRB~170817A even if we introduce the evolution of $p$ with the shock speed by Equation~\eqref{eq.p}, due to the large observational errors. 
The time evolution of $p$ obtained from the calculated spectrum does not much depend on the jet structure, reflecting the fact that the afterglow light curves are similar to each other.

\subsection{Results for an off-axis GRB with a dense ambient medium}\label{sec.result_n0}

\begin{figure*}
	\includegraphics[width=\textwidth]{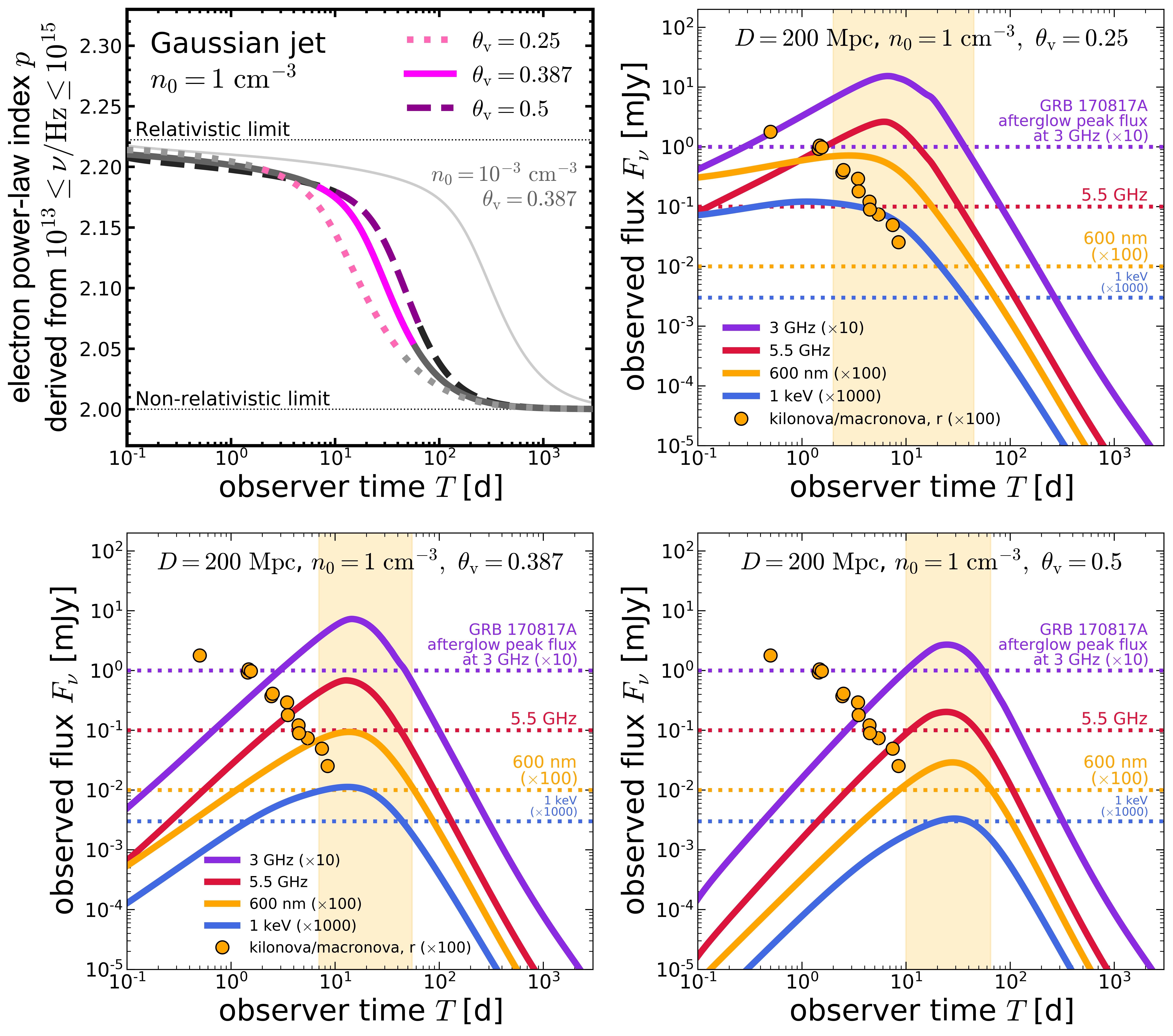}
	\caption{
	{\bf Upper left:}
	Same as the upper-left panel of Figure~\ref{fig.structure_independence} but for $n_0=1$~cm$^{-3}$ and various viewing angles, $\view=0.25$ (dotted line), $\view=0.387$ (bold solid one), and $\view=0.5$ (dashed one), in the case of the Gaussian jet. 
	The coloured section in each line is for the observer time when the optical afterglow flux becomes larger than the peak flux of the GRB~170817A afterglow and an expected optical kilonova/macronova flux, which is indicated by a shade in other panels. 
	For reference, the result for $n_0=1$~cm$^{-3}$ and $\view = 0.387$ (Gaussian jet) in Figure~\ref{fig.structure_independence} is shown by the gray thinner line.
	{\bf The others:}
	Afterglow light curves generated by the Gaussian jet for $D=200$~Mpc, $n_0=1$~cm$^{-3}$, and various viewing angles: $\view=0.25$ ({\bf upper right}), $\view=0.387$ ({\bf lower left}), and $\view=0.5$ ({\bf lower right}).
	For comparison, the peak flux of the GRB~170817A afterglow is shown by a horizontal line with the same colour for each band. 
	The round-shaped points are the expected kilonova/macronova fluxes in {\it r}-band inferred from the kilonova/macronova associated with GRB~170817A, which are obtained by converting the observed kilonova/macronova fluxes \citep{Cowperthwaite17,Drout17} to the values at $D=200$~Mpc.
	The shade in each panel indicates the observer time when the optical afterglow flux is larger than the peak flux of the GRB~170817A afterglow and an optical kilonova/macronova flux.
    }
	\label{fig.denser}
\end{figure*}

Figure~\ref{fig.denser} shows that the evolution of the electron power-law index will be observed in a more luminous afterglow than the GRB~170817A afterglow, if a similar off-axis GRB happens in a dense ambient medium $n_0=1$~cm$^{-3}$ at $D=200$~Mpc.
We here repeated the same calculation of off-axis GRB afterglows as in Section~\ref{sec.result_structure} but with $n_0=1$~cm$^{-3}$ and $D=200$~Mpc.
We used $n_0=1$~cm$^{-3}$ just as a clear illustration that leads to luminous afterglows, while it is the combination of the viewing angle and density that is important for observed afterglow fluxes and, hence, lower densities can be allowed for probing particle acceleration as discussed in Section~\ref{sec.disc.rate}.
$D=200$~Mpc is about the detection horizon of gravitational waves from a neutron star merger for the advanced LIGO in the O4 observational run \citep{LIGOprospects}, while changing the viewing angle $\view$ in a range $0.25 \le \view \le 0.5$, which is about the uncertain range of the viewing angle in GRB~170817A \citep{superluminal}.
The other parameter values remain the same as in Section~\ref{sec.result_structure}.
Hereafter we focus on the Gaussian jet, since our conclusion does not change qualitatively nor quantitatively for the hollow-cone and spindle jets.

The upper-left panel in Figure~\ref{fig.denser} shows the time evolution of the electron power-law index obtained from the calculated spectral slope in $10^{13} \le \nu/\mathrm{Hz} \le 10^{15}$.
The evolutionary paths for $n_0=1$~cm$^{-3}$ are qualitatively the same as for $n_0=10^{-3}$~cm$^{-3}$, but the transition from the relativistic limit to the non-relativistic limit takes place at earlier time, $T\sim10\mathchar`-100$~d, than for $n_0=10^{-3}$~cm$^{-3}$, since the shock is decelerated more efficiently due to the denser ambient medium.
The transition phase corresponds to the peak time of the afterglow light curve for the same reason as for $n_0=10^{-3}$~cm$^{-3}$.
The evolution in the coloured section of each line, where $2.05 \lesssim p \lesssim 2.2$, is observed with larger flux values than the peak flux of the GRB~170817A afterglow and an expected optical kilonova/macronova flux as explained below, indicating the possibility that observational errors become smaller than those for GRB~170817A.

The upper-left panel in Figure~\ref{fig.denser} also reveals that the transition phase tends to be later for a larger viewing angle.
This tendency is due to the fact that the shock speed at the emission centroid for a given observer time is larger for a larger $\view$ since the photons have to leave earlier for the more misaligned observer before the shock is more decelerated, by which the electron power-law index $p$ give by by Equation~\eqref{eq.p} and, hence, the electron power-law index obtained from the calculated spectrum becomes larger for a larger $\view$.
We note that this difference of the photon departure time compensates the difference of the jet energy at the emitting region. 
For example, at $T=10$~d, the brightest point is on the jet axis for $\view=0.25$ while it is still on a jet edge part for $\view=0.5$. Hence, the jet energy at the brightest position is smaller for $\view=0.5$. However, the corresponding laboratory time is earlier for $\view=0.5$ for the reason mentioned above. As a result, the shock Lorentz factor at the emission centroid is still larger for $\view=0.5$ than for $\view=0.25$ at $T=10$~d.

The other panels in Figure~\ref{fig.denser} show afterglow light curves for each viewing angle, whose flux values are to be compared with the peak flux of the GRB~170817A afterglow and an expected kilonova/macronova flux. Due to the large ambient number density $n_0=1$~cm$^{-3}$, the peak flux in each band is larger than that for $n_0=10^{-3}$~cm$^{-3}$ while the peak time is earlier.
As a result, the optical ($600$~nm) flux becomes larger than the peak flux of the GRB~170817A afterglow and an expected kilonova/macronova flux in the observer time indicated by the shade ($2\lesssim T/\mathrm{d} \lesssim 45$ for $\view=0.25$, $7\lesssim T/\mathrm{d} \lesssim 55$ for $\view=0.25$, and $10\lesssim T/\mathrm{d} \lesssim 65$ for $\view=0.5$), which corresponds to the coloured section in the upper-left panel. 
The expected kilonova/macronova fluxes were calculated by rescaling the kilonova/macronova fluxes observed with GRB~170817A to $D=200$~Mpc with the assumption that they do not depend on the viewing angle.

If the viewing angle is decreased, the afterglow becomes more luminous as shown in the three panels in Figure~\ref{fig.denser}, increasing the chance for observations with better quality.
For example, the peak flux at $600$~nm is about two orders of magnitude larger than that in the GRB~170817A afterglow for $\view =0.25$ while it is about one order of magnitude larger for our canonical value $\view =0.387$. 
At the same time, the rising slope becomes shallower and the peak time comes earlier for a smaller viewing angle. 
As a result, the time window for better optical observations (i.e.~the observer time within a shade) spans an earlier phase with a wider range for a more on-axis observer. 
However, we note that the majority in the short GRBs triggered with a gravitational wave signal is expected to have $\view \sim 30^\circ$ \citep{Schutz11}.

We note that X-ray bands would be available for obtaining the electron power-law index from spectra if the effect of the cooling break is appropriately considered:
The spectral slope is $-(p-1)/2$ below the cooling frequency while it is $-p/2$ above it.
It is, hence, important to note that the X-ray flux is also 10 (100) times larger than the peak X-ray flux of the GRB~170817A afterglow at the peak for $\view=0.387$ $(0.25)$.
Radio observations are also useful for deriving the electron power-law index, if the radio frequency is higher than the synchrotron characteristic frequency. 
If available, they help to obtaining the electron power-law index with less errors by increasing the number of observational points spanning a wider frequency range in spectra.
Especially, radio and X-ray fluxes are not contaminated with kilonova/macronova fluxes and, hence, are accessible even in the early time when the optical afterglow is hidden by a kilonova/macronova.

To summarize so far, the time evolution of the electron power-law index could be observed in GRB afterglow spectra with brighter fluxes than the peak flux of the GRB~170817A afterglow and a kilonova/macronova flux, 
if a GRB~170817A-like off-axis GRB occurs in a denser environment with $n_0=1$~cm$^{-3}$ at a typical distance of gravitational events $\sim 200$~Mpc.
A rapid evolution of the electron power-law index from a relativistic value to a non-relativistic one would be observed around the afterglow peak time $T=10-30$~d, which depends on the viewing angle, since the most energetic part of the jet starts to be seen for an off-axis observer as the shock speed slows down to a trans-relativistic speed.
Comparing observations with the theoretical prediction, we will probe particle acceleration at trans-relativistic shocks.

\section{Discussion}\label{sec.discussion}

\subsection{Event rate} \label{sec.disc.rate}
The detectability of the evolution of the electron power-law index is dependent on the measurement accuracy of afterglow spectral slopes. 
In higher density medium, more luminous afterglows could reduce observational errors as discussed in Section~\ref{sec.result_n0}.
Hence, the environment of a short GRB, i.e., the number density of the ambient medium, is a crucial factor, if we assume that the viewing angle and other physical properties are the same as those for GRB~170817A. 
In the following, using the distribution function of the ambient number density in \citet{Berger14} as an example, we roughly estimate that a fraction of $1/2\ \mathchar`- \ 1/10$ (depending on the viewing angle in $0.25 \le \view \le 0.5$) in the GRB~170817A-like off-axis short GRBs at 200~Mpc would give a chance to probe the particle acceleration by observing the evolution of the electron power-law index more accurately than in GRB~170817A.

The afterglow peak flux should exceed that for GRB~170817A at least, if we assume that the error in the electron power-law index obtained from the spectral slope is simply proportional to the afterglow fluxes used for deriving the power-law index.\footnote{We simply ignore a technical and/or strategic discussion on the observational level, while many other factors are involved with errors in estimation of $p$, including the width of the observed frequency range, number of data points in the spectrum, data stacking technique combining observed fluxes obtained on different days, and foreground/background noise, for example.}
The electron power-law index obtained from the calculated spectrum changes $\sim10$ per cent at most ($2\le p \le 2.22$) in our model as shown in the upper-left panel in Figure~\ref{fig.denser}.
Hence, a typical size of errors in the electron power-law index obtained from the spectral slope at each epoch, $|\Delta p|$, has to be smaller than $10$ per cent at least in order to catch the change of $p$.
In the case of GRB~170817A, 
a typical size of errors in the observed electron power-law index is $|\Delta p| = 0.1$ around the afterglow peak time ($T\sim130$~d; See Figure~\ref{fig.structure_independence}), which would roughly correspond to the minimal requirement for our purpose.
For example, \citet{DAvanzo18} derived a photon index corresponding to $p=2.20\pm 0.1$ by using optical and X-ray data around the afterglow peak.

\begin{figure*}
	\includegraphics[width = \textwidth]{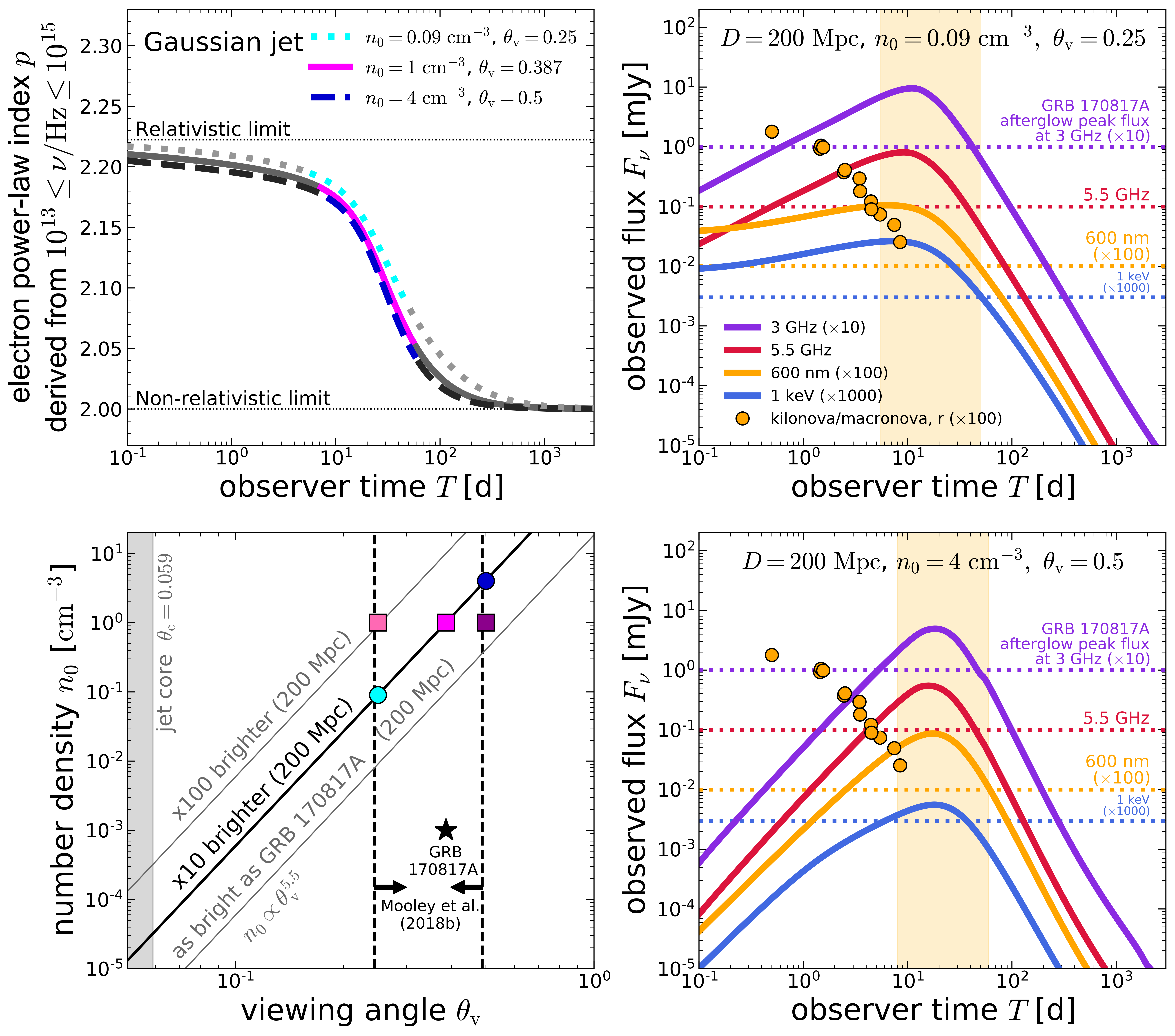}
	\caption{
	{\bf Upper left:}
	Same as the upper-left panel of Figure~\ref{fig.denser} but for various combinations of the number density $n_0$ and viewing angles $\view$. 
	The line for $(n_0, \view) = (1\ \mathrm{cm}^{-3},\ 0.387)$ is the same as in Figure~\ref{fig.denser}.
	{\bf Lower left:}
	Indicated afterglow peak flux at 200~Mpc (for frequencies with $\nu_\mathrm{m} \le \nu \le \nu_\mathrm{c}$) in $(n_0, \view)$ plane. 
	Each diagonal line satisfies $n_0\propto \view^{5.5}$, Equation~\eqref{eq.estimate}.
	The star-shaped point shows our canonical value for GRB~170817A. The square and round points present the values used in Figures~\ref{fig.denser} and \ref{fig.rate}, respectively.
	The vertical dashed lines present the uncertainty of the viewing angle for GRB~170817A  derived from the superluminal motion \citep{superluminal}.
	The gray-shaded region indicates $\view \le \theta_\mathrm{c}$, where $\theta_\mathrm{c}=0.059$ is the standard deviation of the Gaussian jet structure.
	{\bf Right panels:}
	Same as the right panels in Figure~\ref{fig.denser} but for 
	various combinations of the number density and viewing angles: 
	$(n_0, \view) = (0.09\ \mathrm{cm}^{-3},\ 0.25)$ ({\bf upper}) and 
	$(n_0, \view) = (   4\ \mathrm{cm}^{-3},\ 0.5)$ ({\bf lower}).
	}
	\label{fig.rate}
\end{figure*}

The afterglow peak flux for an off-axis observer generally obeys the following scaling law \citep{Xie18,GNP19}:
\begin{align}
    F_{\nu,\mathrm{p}} \propto n_0^{(p+1)/4}\view^{-2p} D^{-2},
    \label{eq.scaling}
\end{align}
for fixed jet energy, $\nu$, $\ee$, $\eB$, and $p$, where the observed frequency is assumed to satisfy $\nu_\mathrm{m} \le \nu \le \nu_\mathrm{c}$.
Substituting $p=2.17$, which is a representative value of the electron power-law index around the peak time in our model (see Figures~\ref{fig.structure_independence} and \ref{fig.denser}), into Equation~\eqref{eq.scaling}, we obtain
\begin{align}
    n_0 &= 1 \ \mathrm{cm}^{-3}
        \left(\frac{F_{\nu,\mathrm{p}}}{10F_{\nu,\mathrm{p},\mathrm{170817A}}}\right)^{1.3}
        \left(\frac{\view}{0.387}\right)^{5.5}
        \left(\frac{D}{200\ \mathrm{Mpc}}\right)^{2.5},
        \label{eq.estimate}
\end{align}
with the canonical values for GRB~170817A, $n_{0,\mathrm{170817A}}=10^{-3}$~cm$^{-3}$, $\theta_{\mathrm{v},\mathrm{170817A}} = 0.387$, and $D_\mathrm{170817A}=41$~Mpc.
The relation between $n_0$ and $\view$ given by Equation~\eqref{eq.estimate} is shown in the lower-left panel in Figure~\ref{fig.rate} by each diagonal line.

We demand, for example, the peak flux that is 10 times larger than that for GRB~170817A so that afterglow fluxes in some period exceed the peak flux of the GRB~170817A afterglow. 
Then, the allowed area in $(n_0,\view)$ plain is above the thick diagonal line in the lower-left panel of Figure~\ref{fig.rate}. 
Equation~\eqref{eq.estimate} holds, in fact, on the line with $0.25 \le \view \le 0.5$ (the round-shaped points, for example) as demonstrated in the right panels in Figure~\ref{fig.rate} (see the optical peak flux). 
There is a time window when the optical afterglow is more luminous than the peak afterglow flux of GRB~170817A (the shaded region). 
In the period, an off-axis observer could more accurately detect the evolution of the electron power-law index, which is presented in the upper-left panel of Figure~\ref{fig.rate}.
We note that the evolution of the electron power-law index obtained from the spectral slope and the observable time window does not much depend on the combination of $(n_0,\view)$ on a diagonal line at least for $0.25 \le \view \le 0.5$. 

The number density should satisfy $n_0 > 0.09\ \mathrm{cm}^{-3}$ for $\view =0.25$, $n_0 > 1\ \mathrm{cm}^{-3}$ for $\view =0.387$, or $n_0 > 4\ \mathrm{cm}^{-3}$ for $\view =0.5$ as seen in the lower-left panel of Figure~\ref{fig.rate}.
The distribution of the number density of the ambient medium for short GRBs has been studied observationally and theoretically \citep{Belc06,Berger14,Fong15,Wiggins18,OConnor20},
although it depends on the limited sample of short GRBs, uncertain parameters used for estimate, and/or a model of stellar evolution and mass distribution in galaxies.
Using the distribution function of figure~17 in \citet{Berger14} as an example, 
we find that the chance probabilities for $n_0 > 0.09\ \mathrm{cm}^{-3},\ 1\ \mathrm{cm}^{-3},\ 4\ \mathrm{cm}^{-3}$ are $\sim0.5,\ 0.3,\ 0.1$, respectively. In other words, particle acceleration at trans-relativistic shocks would be probed in about one in several or ten of short GRB afterglows at 200~Mpc, which will be found as a counterpart of gravitational wave signals, if the physical properties and viewing angle are the same as GRB~170817A.

We note that 
the event rate can increase if another particle acceleration model holds and $p$ changes in a wider range of value, which alleviates the demand on observational errors for observing the evolution of $p$.
For instance, the model B in figure 2 of \citet{Aoi08} shows that the value of $p$ drops by $\sim 25$ per cent from the relativistic to non-relativistic limits.
We note, however, that the evolution of $p$ in theory has diversity, since there is a variety of particle acceleration models. For example, figure~2 in \citet{Aoi08} shows that the value of $p$ could increases with shock deceleration in some model and figure~2 in \citet{NK21} shows that the evolution of $p$ can be non-monotonic for the shock 4-velocity. Even for non-relativistic shocks, the value of $p$ can be different from a standard value of $p=2$, depending on the anisotropy of accelerated particles \citep{TK15} and/or whether diffusion is regular or not \citep{Kirk96,PZ12,LY14}.

A wide range of $p$ measured in GRBs and SNRs also suggests that our estimate of the event rate is rather conservative, although the range is not obtained from a single shock evolution but from many objects with different shocks:
The value of $p$ measured in other GRBs ranges from $\sim 1$ to $\sim 3$ \citep{PK02,Shen06,Starling08,Curran09,Curran10,Fong15,Gompertz18,Troja19}.
$p\sim 3$ is also observed in SNRs with $\beta_\mathrm{sh}\sim 0.1$, while $p\sim 2$ is observed in older SNRs with a slower shock speed $\beta_\mathrm{sh}\sim 0.01$ \citep{Bell11,Bozzetto17}.
The origin of these values has not been understood well and is beyond the scope of this paper.
In any case, it will be useful to probe particle acceleration at trans-relativistic shocks for understanding these observations.

We also note that the observable event rate can increase in the future as detectors improve their sensitivity, of course. For instance, {\it Athena} is a future mission that has about $10\mathchar`-100$ times better sensitivity than the current X-ray facilities and is expected to probe trans-relativistic particle acceleration in dimmer off-axis GRB afterglows \citep{Athena}.

\subsection{Effects of sideways expansion of the jet on the spectral evolution}\label{sec.sideways}
Even if we take into account the sideways expansion of the jet, the result in Section~\ref{sec.result_n0} does not qualitatively change. That is, the electron power-law index obtained from the calculated spectrum transits the trans-relativistic value at around the afterglow peak time and it will be observed with more luminous afterglow fluxes.
We simply ignored the sideways expansion of the jet so far, since
the sideways expansion is a very slow process where the jet angle size increases logarithmic in time, which starts when the shock Lorentz factor is sufficiently reduced \citep{Cannizzo,ZM09,Wygoda,DeColle,EM12}.
However, the lateral spreading of the jet quantitatively alters the result.
Below we show the quantitative difference between with and without the lateral spreading by incorporating the sideways expansion of the jet in our model.

In the case of structured jets, the sideways expansion is introduced by the prescription of concentric top-hat jets \citep{Ryan,FKL}.
It is a rather provisional prescription that treats a structured jet as a superposion of top-hat jets with different equivalent isotropic energies and opening angles and assumes that each top-hat jet independently expands sideways.
One prescription that models the spreading behaviour of top-hat jets resonably well in practice is the one given by \citep{GP12,FKL}:
\begin{align}
    \label{eq.expansionRate}
    \frac{\diff \theta_{\mathrm{j},n}}{\diff \ln R_n} = \frac{1}{\Gamma_{\mathrm{sh},n}^2\theta_{\mathrm{j},n}}.
\end{align}
In the above equation, $\theta_{\mathrm{j},n}$ is the jet opening angle of the $n$-th top-hat jet. $R_n$ and $\Gamma_{\mathrm{sh},n}$ are the shock radius and the shock Lorentz factor of the $n$-th top-hat jet, respectively,
where a structured jet is divided into $N$ top-hat jets and $n=1,\cdots,N$ is the label of a top-hat jet ($\theta_{\mathrm{j},1}<\cdots<\theta_{\mathrm{j},N}$).
Each top-hat jet is assumed to expand sideways when the shock speed is slower than $\Gamma_\sh = \Gamma_{\sh,\mathrm{max}}=100$.
The jet energy contained in each top-hat jet is assumed to be conserved during sideways expansion and, as a result, the isotropic equivalent energy of the $n$-th jet $E_{\mathrm{iso},n}(t)$ becomes smaller as the jet spreads laterally.
We put $\diff \theta_{\mathrm{j},n}/\diff t =0$ when $E_{\mathrm{iso},n}(t) > E_{\mathrm{iso},n+1}(t)$, otherwise an inner top-hat jet can overtake outer ones in our calculation.
We note that this prescription will overestimate sideways expansion, since the expansion of each part of a structured jet would not be independent in reality but would be suppressed by the outer part of the jet while we impose a minimal requirement that an inner top-hat jet does not overtake outer ones.
Hence, a real jet dynamics will lie between the dynamics with and without the lateral spreading with this prescription.
In this sense, we could put an upper limits on the effect of sideways expansion with the above prescription.

\begin{figure*}
	\includegraphics[width = \textwidth]{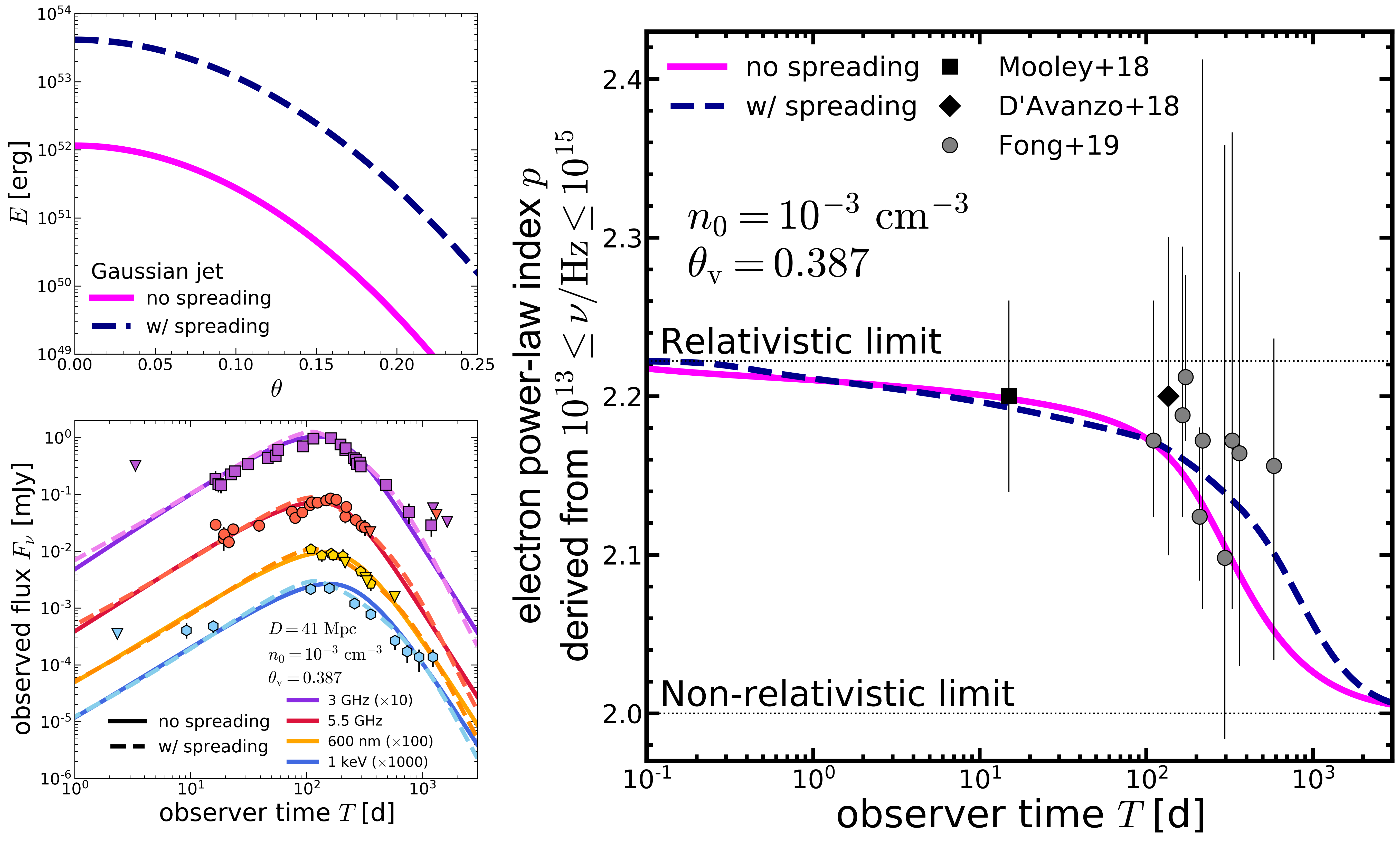}
	\caption{
	{\bf Upper left:} Gaussian jet structures that are consistent with the afterglow of GRB~170817A for $n_0=10^{-3}$~cm$^{-3}$, $\view=0.387$, and $\ee = 0.1$.
	The solid and dashed lines are for the cases without and with lateral spreading of the jet, respectively.
	The magenta solid line is the same as in Figure~\ref{fig.jet}.
	{\bf Lower left:} Afterglow light curves that are produced by the Gaussian jets in the upper-left panel.
	The solid and dashed lines are for the cases without and with lateral spreading, respectively, where
	the value of $\eB$ is set to $2.44\times 10^{-4}$ and $2.44\times 10^{-6}$ for these respective cases.
	The meaning of the colors and the observational data points are the same as in the left panel of Figure~\ref{fig.Gaussian_fid}.
	{\bf Right:} 
	Evolution of the electron power-law index obtained from the calculated afterglow spectrum. 
	The solid and dashed curves are for the cases without and with lateral spreading, respectively.
	The magenta solid line and the data points are the same as in the left-top panel of Figure~\ref{fig.structure_independence}.
	}
	\label{fig.spreading}
\end{figure*}

The upper-left panel of Figure~\ref{fig.spreading} shows a Gaussian jet structure that is calibrated by the afterglow of GRB~170817A with the sideways expansion (dashed line). 
The Gaussian jet produces consistent afterglow light curves (dashed lines in the lower-left panel)
with the same parameter values as those used in Section~\ref{sec.result_structure} except for $\eB$.
The jet energy is increased in order to adjust the peak time of the afterglow, since the shock wave slows down earlier with the lateral spreading. Accordingly, the value of $\eB$ is decreased as $\eB=2.44\times 10^{-6}$ for adjusting the afterglow flux.
We note that a true jet energy and value of $\eB$ will lie between those with and without sideways expansion, since we assumed the maximal expansion rate as mentioned above.
The right panel of Figure~\ref{fig.spreading} shows that the evolution of the electron power-law index obtained from the calculated spectrum with lateral spreading is also consistent with the observations within observational errors.
We note again that a true evolutionary curve will lie between the cases with and without sideways expansion.

\begin{figure*}
	\includegraphics[width = \textwidth]{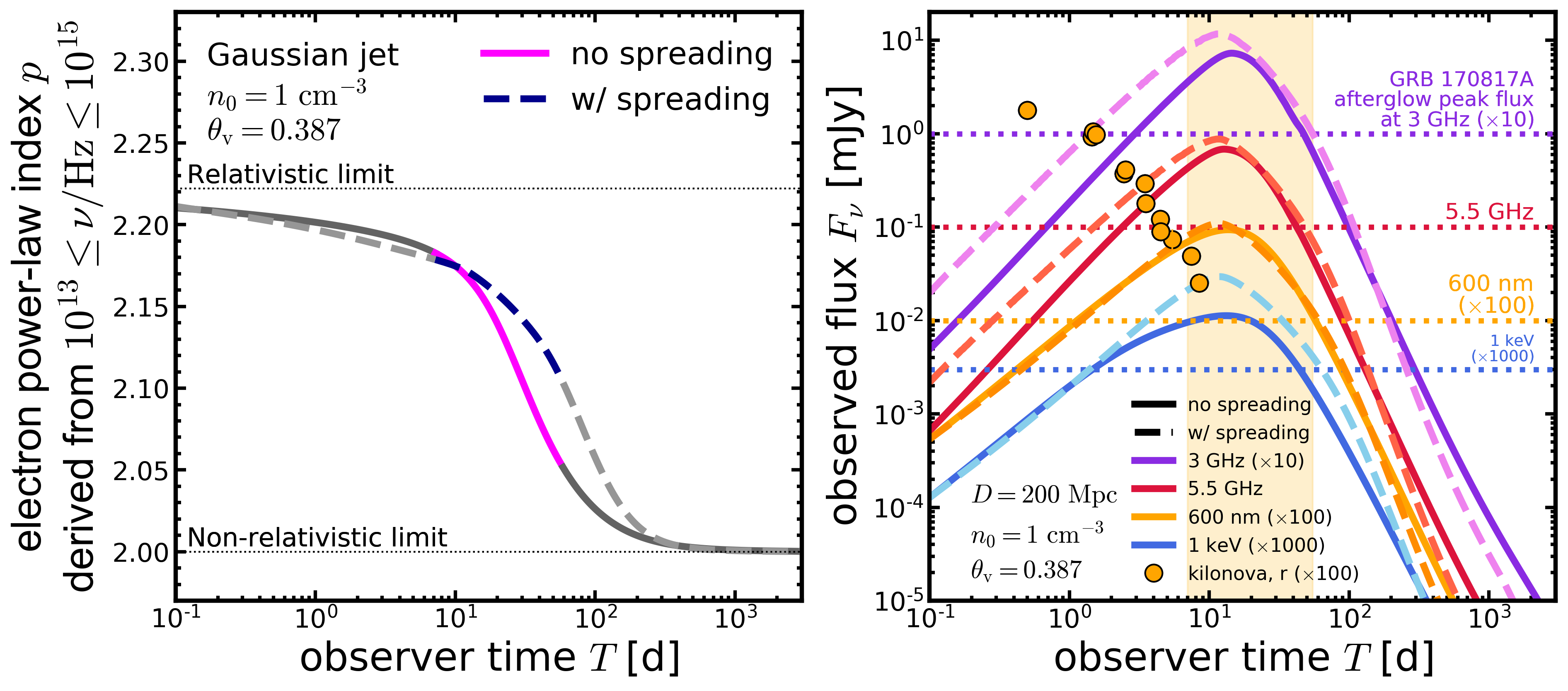}
	\caption{
	{\bf Left:} Same as the right panel of Figure~\ref{fig.spreading} but for a denser ambient medium with $n_0=1$~cm$^{-3}$.
	The coloured section in each line is for the observer time when the optical afterglow flux becomes larger than the peak flux of the GRB~170817A afterglow and an optical kilonova/macronova flux, which is indicated by a shade in the right panels.
	The solid line is the same as that in Figure~\ref{fig.denser}.
	{\bf Right:} Afterglow light curves generated by the Gaussian jets for $D=200$~Mpc, $n_0=1$~cm$^{-3}$, $\view=0.387$, and $\ee = 0.1$. The value of $\eB$ is set to $2.44\times 10^{-4}$ and $2.44\times 10^{-6}$ for the cases without and with lateral spreading, respectively. 
	The shade indicates the observer time when the optical (600~nm) afterglow flux is larger than the peak flux of the GRB~170817A afterglow and an optical kilonova/macronova flux.
	The solid lines, horizontal dotted lines, and kilonova/macronova fluxes are the same as those in the lower-left panel of Figure~\ref{fig.denser}.
	}
	\label{fig.spreading_prediction}
\end{figure*}

Figure~\ref{fig.spreading_prediction} shows the evolution of the electron power-law index obtained from the calculated spectrum and afterglow light curves for a higher number density $n_0=1$~cm$^{-3}$ with sideways expansion.
A real evolutionary curve of the electron power-law index obtained from the calculated spectrum will lie between the cases with and without lateral spreading.
The colored segment is used for the observer time when the optical afterglow flux becomes more luminous than the peak flux of the GRB~170817A afterglow and an expected kilonova/macronova flux, which is almost identical for the cases with and without sideways expansion.
Hence, we come to the same conclusion that the change of the electron power-law index could be more accurately measured than for GRB~170817A.
The estimate of the event rate in Section~\ref{sec.disc.rate} would not change, since the optical light curves for the cases with and without lateral expansion are quantitatively the same.

\section{Summary \& Conclusions}\label{sec.conclusion}
We studied a potential use of electromagnetic counterparts to gravitational waves, off-axis GRB afterglows, for probing particle acceleration at trans-relativistic shocks by calculating afterglow light curves and spectra with a particle acceleration model.
As a first step, we used a particle acceleration model of \citet{KW05}, where the electron power-law index $p$ changes with the shock speed as given by Equation~\eqref{eq.p}.
Particle acceleration could be probed by comparing the time evolution of $p$ obtained from the calculated spectrum and observed one.
We showed that this is promising since a rapid change of $p$ is predicted around the afterglow peak time with fluxes larger than the peak flux of the GRB~170817A afterglow, if a GRB~170817A-like off-axis GRB takes place in a denser environment
at a typical luminosity distance of gravitational wave events $D=200$~Mpc.

First, in Section~\ref{sec.result_structure},
we found that the observed spectral evolution in the GRB~170817A afterglow is also consistent with our model with evolving $p$ within the observational errors, while models with a constant $p$ are used for GRB~170817A in the literature.
The time evolution of $p$ obtained from the calculated spectrum does not much depend on the jet structure, reflecting the similarity of the afterglow light curves.

Second, in Section~\ref{sec.result_n0},
we showed that particle acceleration could be probed by future GRB~170817A-like off-axis GRBs in a denser environment with $n_0=1$~cm$^{-3}$ at a typical distance of gravitational wave counterparts.
A rapid transition of $p$ from a relativistic value to a non-relativistic one would be observed in a more luminous afterglow than the peak flux of the GRB~170817A and a kilonova/macronova flux.
The rapid evolution of $p$ takes place around the afterglow peak, $T=10\char`-30$~d, which depends on the viewing angle.
Larger afterglow fluxes increase the chance for observing the time evolution of the power-law index with less errors than those in GRB~170817A.

We roughly estimated that $10\char`-50$ per cent, depending on the viewing angle, of the GRB~170817A-like off-axis short GRBs at 200~Mpc has higher afterglow fluxes than the peak flux of the GRB~170817A and give a chance to probe the particle acceleration as discussed in Section~\ref{sec.disc.rate}, whereas our estimate could suffer from the uncertainty in the distribution function of the ambient number density. Our conclusions and estimate above do not change even if we take into account the sideways expansion of the jet as discussed in Section~\ref{sec.sideways}.

As shown in this paper, off-axis GRBs are one of the unique astrophysical sources that could probe particle acceleration at trans-relativistic shocks.
Such GRBs in a dense environment would be observed with gravitational wave signals in the upcoming observational runs of LIGO/Vigo.
While we demonstrated the use of off-axis GRB afterglows by using a particle acceleration model,
a comprehensive study will be necessary for comparison with future observations.
It is future work to consider other particle acceleration models taking into account, for example, 
perpendicular shocks \citep{TK15,Kamijima20}, anisotropic diffusion \citep{Keshet06,Keshet20}, turbulent field in upstream and/or downstream flows \citep{NO04,NO06,Niemiec06,Kamijima20}, multidimensional effects \citep{Keshet17,Lavi20}, particle feedback to a diffusion function \citep{NK21}, and large-angle scattering \citep{Aoi08,Keshet20}. It is also a future work to consider the second order Fermi acceleration by downstream turbulence, which could happen after particles are accelerated by the first order Fermi acceleration around the shock front and could modify the spectrum \citep{AT09, Ohira13, Pohl15, YO20}. The downstream turbulence is driven by a shock wave propagating into non-uniform medium \citep{SG07,Inoue11,Tomita22} and the spectral modification depends on the upstream inhomogeneity \citep{YO20}. Hence, a precise measurement of the GRB afterglow spectrum may be also useful for probing the GRB environment as well as the particle acceleration process, which will be studied in a forthcoming paper.

\section*{Acknowledgements}
We thank the YITP workshops YITP-T-21-05, YITP-T-19-04, and YKIS2019.
K. T. thanks Dr.~Koutarou Kyutoku for useful comments and code comparison.
K.T. also thanks Drs.~Susumu Inoue, Ryo Yamazaki, and Johann Cohen-Tanugi for useful comments.
K. T. and K. I. thank Drs.~Hamid Hamidani, Wataru Ishizaki, and Tomoki Wada for daily discussion.
H. J. van Eerten acknowledges partial support by the European Union Horizon 2020 Programme under the AHEAD2020 project (grant agreement number 871158).
This work is supported by JSPS Grants-in-Aid for Scientific Research
17H06362 (K. T., K. I.),
22H00130, 20H01904, 20H01901, 20H00158, 18H01215, 17H06357, 17H06131 (K. I.),
and 19H01893, 21H04487 (Y. O.).

\section*{Data Availability}
The data underlying this article will be shared on reasonable request
to the corresponding author.








\appendix
\section{Relativistic and non-relativistic limits for downstream quantities}\label{app.downstream}

In the relativistic limit ($\Gamma \gg 1$), Equations~(\ref{eq.RH1})-(\ref{eq.zeta_Gamma}) yield the well-known result for the relativistic strong shock:
\begin{align}
    \hat{\gamma}' &= \frac{4}{3},\\
    \label{eq.strong_rela_n}
    n' &= 4\Gamma n_0, \\
    \label{eq.strong_rela_e}
    e_\mathrm{i}' &= \Gamma n'\mpr c^2 = 4\Gamma^2n_0\mpr c^2, \\
    \label{eq.strong_rela_Gamma}
    \Gamma &= \frac{1}{\sqrt{2}}\Gamma_\sh.
\end{align}
In the non-relativistic limit, on the other hand, Equations~(\ref{eq.RH1})-(\ref{eq.zeta_Gamma}) recover the strong shock limit in the Newtonian phase:
\begin{align}
    \hat{\gamma}' &= \frac{5}{3},\\
    n' &= 4n_0, \\
    \label{eq.strong_NR_e}
    e_\mathrm{i}' &= \frac{1}{2} n'\mpr \beta^2c^2, \\
    \label{eq.strong_NR_beta}
    \beta &= \frac{3}{4}\beta_\sh.
\end{align}

\section{Derivation of Equation~(29)}\label{app.eq}

\begin{figure*}
    \centering
    \begin{minipage}[t]{.49\hsize}
        \centering
	    \includegraphics[width = \columnwidth]{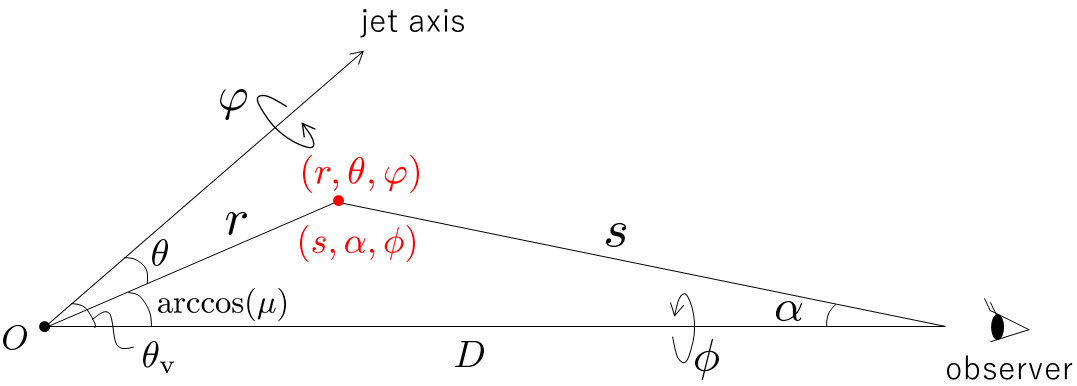}
	\end{minipage}
	\begin{minipage}[t]{.49\hsize}
	    \centering
	    \includegraphics[width = \columnwidth]{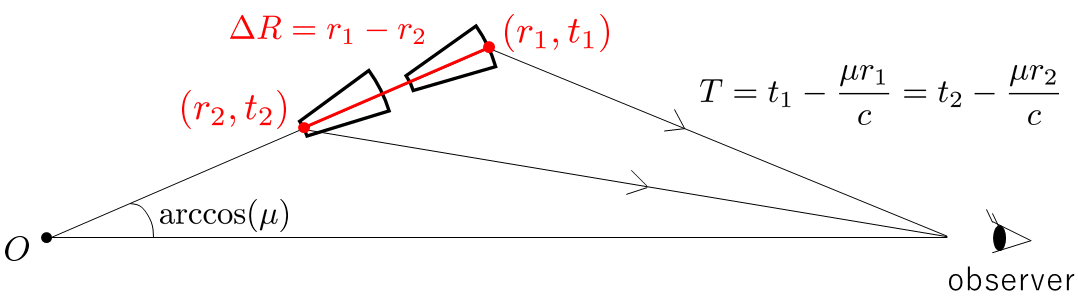}
	\end{minipage}
	\caption{
    {\bf Left:}
    Geometrical illustration for calculating the observed flux.
    $(r,\theta,\varphi)$ is the spherical coordinates whose origin is at the explosion site of a GRB (denoted by $O$), while $(s,\alpha,\phi)$ is the spherical coordinates whose origin is at the observer at a distance $D$ from the explosion site. $\view$ is the viewing angle measured from the jet axis. $\mu$ is the cosine of the angle spanned by a radial direction and the line of sight, which is given by Equation~\eqref{eq.mu}. 
    {\bf Right:}
    Geometrical illustration for the calculation in the optically thin limit.
    $\Delta R$ shows the region where emitted photons arrive at the observer at $T$, which is given by Equation~\eqref{eq.deltaR}. Two fan-shaped objects drawn by thick black lines show a part of the shocked region at different laboratory times. The head and tail of each object corresponds to the shock front and the tail of the shocked region, respectively.}
	\label{fig.geometry}
\end{figure*}

This Section gives a derivation of Equation~\eqref{eq.Fnu}. We note that Equation~\eqref{eq.Fnu} is superior to eq.~(A20) in \citet{Eerten10} in terms of taking into account the difference of the shock radius at different angles in Equation~\eqref{eq.dadmu1}.

The basic equations for calculating the observed flux are given by \citep{RL}
\begin{align}
    \label{eq.transf}
    \frac{\diff I_\nu}{\diff s} &= -\alpha_\nu I_\nu + j_\nu,\\
    \label{eq.basic}
    F_\nu &= \int \diff \alpha \sin \alpha \int \diff \phi I _\nu \cos \alpha.
\end{align}
The upper equation is the transfer equation, where $I_\nu$ is the intensity of a ray with $\nu$ being the frequency, $s$ is the length along the ray, $\alpha_\nu$ is the absorption coefficient, and $j_\nu$ is the emission coefficient.
The lower equation gives the observed flux, which integrates the rays passing through the observer, where $\alpha$ is the polar angle measured from the line of sight and $\phi$ is the azimuthal angle around the line of sight (See Figure~\ref{fig.geometry}).
Hereafter, we suppose that the observer lies at a luminosity distance $D$ in the direction of $(\theta,\varphi)=(\view,0)$, where $\view$ is the viewing angle. 
The cosine of the angle between a radial direction and the line of sight, $\mu$, is then given by Equation~\eqref{eq.mu}.

In Appendix~\ref{app.thin}, we consider the optically thin limit of Equations~\eqref{eq.transf} and \eqref{eq.basic}. In Appendix~\ref{app.abs}, we consider a general case with absorption and derive Equation~\eqref{eq.Fnu} with the aid of the result given in Appendix~\ref{app.thin}. Appendix~\ref{app.s} gives an interpretation of the result in Appendix~\ref{app.abs}.

\subsection{Optically thin limit}\label{app.thin}
First, we review the case of the optically thin limit with $\alpha_\nu = 0$.
In this case, Equations~\eqref{eq.transf} and \eqref{eq.basic} give
\begin{align}
    \label{eq.Fnu_1_thin}
    F_\nu = \int \diff \alpha \sin \alpha \int \diff \phi \int \diff s j_\nu \cos \alpha.
\end{align}
We now change the integral variables from $(s, \alpha, \phi)$ to $(r, \theta, \varphi)$ by calculating the Jacobian:
\begin{align}
    \diff s \diff \alpha \diff \phi &= 
        \left|\frac{\partial (s, \alpha, \phi )}{\partial(r, \mu, \phi)}\right|
        \left|\frac{\partial (r, \mu, \phi )}{\partial(r, \theta, \varphi)}\right|\diff r \diff \theta \diff \varphi,\\
        &=\left|\frac{\partial s}{\partial r} \frac{\partial \alpha}{\partial \mu} - \frac{\partial s}{\partial \mu} \frac{\partial \alpha}{\partial r} \right| \sin \theta \diff r \diff \theta \diff \varphi,\\
        \label{eq.Jacobian_thin}
        &\sim \frac{r\sin \theta}{D\sqrt{1-\mu^2}},
\end{align}
where we used the following relations for calculating derivatives:
\begin{align}
    s \sin \alpha &= r \sqrt{1-\mu^2},\\
    s \cos \alpha &= D - r\mu,
\end{align}
and took the limit of $r \ll D$ in the third line, where $D$ is the luminosity distance to the origin, $O$. In the limit of $r \ll D$, we also find
\begin{align}
    \label{eq.sin_thin}
    \sin \alpha &\sim \alpha \sim \frac{r\sqrt{1-\mu^2}}{D},\\
    \label{eq.cos_thin}
    \cos \alpha &\sim 1.
\end{align}
Substituting Equations~\eqref{eq.Jacobian_thin}, \eqref{eq.sin_thin}, and \eqref{eq.cos_thin} into Equation~\eqref{eq.Fnu_1_thin}, we obtain
\begin{align}
    \label{eq.Fnu_2_thin}
    F_\nu \sim \frac{1}{D^2}\int \diff \theta \sin \theta \int \diff \varphi \int \diff r r^2 j_\nu.
\end{align}

We here approximate the integral with respect to $r$ as follows.
The observed flux at a given observer time $T$ is only contributed from a limited region on each radial integral path, $\Delta R = r_1 - r_2$, since the emission of photons take place only in the shocked region (See Figure~\ref{fig.geometry}). 
Here, $r_1$ corresponds the radius of the shock front at $t_1$, 
\begin{align}
    \label{eq.r_1_thin}
    r_1 = R(t_1,\theta),
\end{align}
while
$r_2$ corresponds the radius of the tail of the shocked region at $t_2$,
\begin{align}
    \label{eq.r_2_thin}
    r_2 = R(t_2,\theta) - w(t_2,\theta),
\end{align}
where $R(t, \theta)$ is the shock radius given by Equation~\eqref{eq.R} and $w$ is the width of the shocked region. The laboratory time $t_{1,2}$ satisfies the condition that emitted photons arrive at $T$, which is given by
\begin{align}
    \label{eq.samearrivaltime}
    T = t_1 - \frac{\mu r_1}{c} = t_2 - \frac{\mu r_2}{c},
\end{align}
where
we chose $T=0$ as the arrival time of a photon emitted at the origin, $O$, at $t=0$.
We assume that the shock speed is nearly constant on the journey from $r_1$ to $r_2$, by which $R(t_1,\theta)$ is given by
\begin{align}
    \label{eq.relationbtwR1andR2_thin}
    R(t_1,\theta) \sim R(t_2,\theta) + c\beta_\sh(t_1 - t_2).
\end{align}
The width of the shocked region, $w$, is given by the conservation of the swept up mass \citep{Granot99,Eerten10}. Assuming that the shell width $w$ is thin, we obtain
\begin{align}
    &\frac{\Omega}{3}R^3 n_0 = \Omega R^2 w 4\Gamma^2 n_0,\\
    \label{eq.w}
    \Rightarrow\quad & w = \frac{R}{12\Gamma^2},
\end{align}
where $\Omega$ is the solid angle of the considered section on the jet surface and $4\Gamma^2 n_0$ is the density at the shock downstream in the relativistic limit, cf.~Equation~\eqref{eq.strong_rela_n}.
Equations~\eqref{eq.r_1_thin}-\eqref{eq.relationbtwR1andR2_thin} and \eqref{eq.w} give $\Delta R$ as follows \citep{Eerten10}:
\begin{align}
    \label{eq.deltaR}
    \Delta R = r_1 - r_2 = \frac{R(t_2,\theta)}{12\Gamma^2(1-\mu \beta_\sh)}.
\end{align}
By using $\Delta R$ in Equation~\eqref{eq.deltaR} and the assumption that $j_\nu$ is constant on $\Delta R$, we approximate Equation~\eqref{eq.Fnu_2_thin} as \citepalias{TI20,TI21}
\begin{align}
    F_\nu &\sim \left. \frac{1}{D^2}\int \diff \theta \sin \theta \int \diff \varphi  \Delta R R^2 j_\nu \right|_{t=T+\mu R/c} ,\\
    \label{eq.Fnu_3_thin}
    &=\left. \frac{1}{D^2}\int \diff \theta \sin \theta \int \diff \varphi  \frac{R^3}{12\Gamma^2(1-\mu \beta_\sh)} j_\nu \right|_{t=T+\mu R/c}.
\end{align}

\begin{figure*}
    \centering
    \begin{minipage}[t]{.49\hsize}
        \centering
	    \includegraphics[width = \columnwidth]{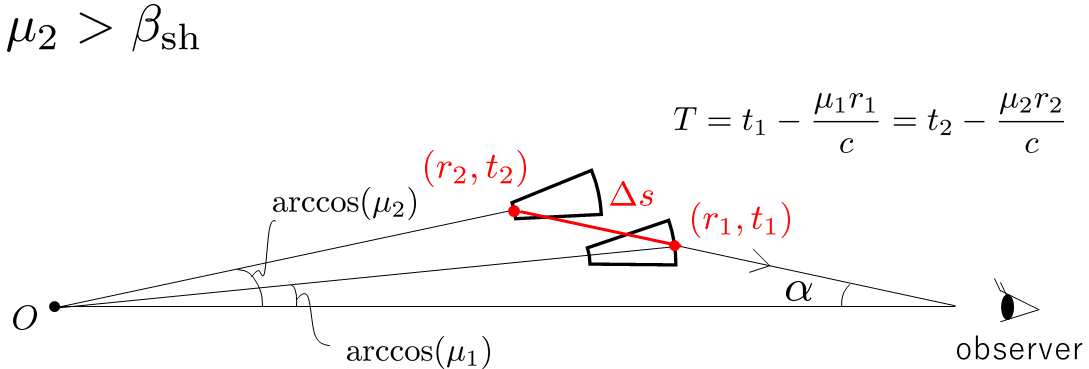}
	\end{minipage}
	\begin{minipage}[t]{.49\hsize}
	    \centering
	    \includegraphics[width = \columnwidth]{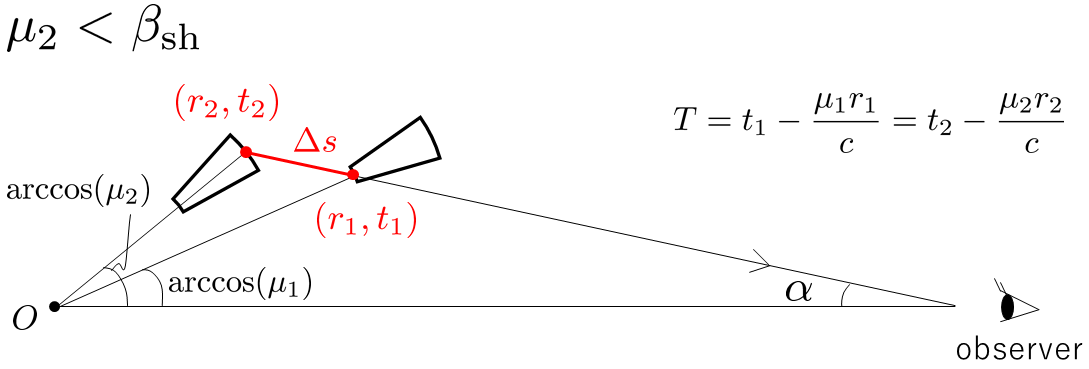}
	\end{minipage}
	\caption{
	Illustration of $\Delta s$ given by Equation~\eqref{eq.s}, where $\Delta s$ gives the length on a ray where emitted photons arrive at the observer at $T$.
    The left and right panels show the cases with $\mu_2 > \beta_\sh$ and $\mu_2 < \beta_\sh$, respectively.
    }
    \label{fig.geometry2}
\end{figure*}

\subsection{General case with absorption}\label{app.abs}
We here consider a general case with absorption ($\alpha _\nu \ne 0$).
In our purpose,
the intensity on a given ray is contributed from a limited length on the ray, $\Delta s$, since the emission and absorption of photons take place only in the shocked region.
Assuming that $\alpha_\nu$ and $j_\nu$ are constant along a ray in the shocked region, the intensity of a ray can be given by physical quantities at the shock front on the ray. 
Indeed, we can solve the transfer equation, Equation~\eqref{eq.transf}, as
\begin{align}
    \label{eq.Inu}
    I_\nu = \frac{j_\nu}{\alpha _\nu} (1 - e^{-\tau_\nu}),
\end{align}
where
\begin{align}
    \label{eq.opacity}
    \tau _\nu = \int \alpha_\nu \diff s = \alpha_\nu \Delta s
\end{align}
is the opacity and $\Delta s$ is given later by Equation~\eqref{eq.s}. 
Substituting Equation~\eqref{eq.Inu} into Equation~\eqref{eq.basic}, we obtain
\begin{align}
    \label{eq.Fnu_1}
    F_\nu (T) = \left. \int \diff \alpha \sin \alpha \int \diff \phi \frac{j_\nu}{\alpha _\nu} (1 - e^{-\tau_\nu}) \cos \alpha \right|_{t=T+\mu R/c},
\end{align}
where the integrand is evaluated at the shock radius at the laboratory time when the emitted photons reach the observer at an observer time $T$:
\begin{align}
    \label{eq.equivt}
    t = T + \frac{\mu R}{c}.
\end{align}

Now, we change the integral variables in Equation~\eqref{eq.Fnu_1} from $(\alpha,\phi)$ to $(\theta,\varphi)$ by calculating the Jacobian:
\begin{align}
    \diff \alpha \diff \phi &= \left|\frac{\partial (\alpha, \phi )}{\partial(\mu, \phi)}\right|
        \left|\frac{\partial (\mu, \phi )}{\partial(\theta, \varphi)}\right|\diff \theta \diff \varphi,\\
        \label{eq.Jacobian}
        &=\left|\frac{\partial \alpha}{\partial \mu}\right| \sin \theta \diff \theta \diff \varphi.
\end{align}
We note here that
\begin{align}
    \label{eq.sin}
    \sin \alpha &\sim \alpha \sim \frac{R\sqrt{1-\mu^2}}{D},\\
    \label{eq.cos}
    \cos \alpha &\sim 1,
\end{align}
in the limit of $R \ll D$. Note that $R=R(t,\theta)$ in Equation~\eqref{eq.sin} is not a radial coordinate but is the shock radius, which depends on angular coordinates through the jet structure and the laboratory time $t(\mu)$ that satisfies Equation~\eqref{eq.equivt}.
Using Equation~\eqref{eq.sin}, we obtain $\partial \alpha/\partial \mu$ in Equation~\eqref{eq.Jacobian} as follows:\footnote{The first term in Equation~\eqref{eq.dadmu1} was simply omitted in \citet{Eerten10}.}
\begin{align}
    \label{eq.dadmu1}
    \frac{\partial \alpha}{\partial \mu} \sim \frac{\partial R}{\partial \mu}\frac{\sqrt{1-\mu^2}}{D} - \frac{\mu R}{D\sqrt{1-\mu^2}}.
\end{align}
$\partial R/\partial \mu$ is obtained from Equations~\eqref{eq.R} and \eqref{eq.equivt} after differentiating them with respect to $\mu$ and eliminating $\partial t/\partial \mu$ as follows:
\begin{align}
    \frac{\partial R}{\partial \mu} &= \frac{1}{1-\mu\beta_\sh}\left(\beta_\sh R + c\int_0^t \frac{\partial \beta_\sh}{\partial \mu} \diff t\right),\\
    \label{eq.dRdmu_middle}
    &= \frac{\beta_\sh R(1+\lambda)}{1-\mu\beta_\sh},\\
    \label{eq.dRdmu}
    &\sim \frac{\beta_\sh R}{1-\mu\beta_\sh},
\end{align}
where the first and second terms in the top equation come from the difference of the laboratory time and the jet structure, respectively. $\lambda$ in the middle equation is defined by
\begin{align}
    \label{eq.chi}
    \lambda \equiv \frac{c}{\beta_\sh R}\int _0^t \frac{\partial \beta_\sh}{\partial \mu} \diff t,
\end{align}
which was neglected in the bottom equation, 
since we find $|\lambda| \lesssim 1$ in the luminous region on the jet surface that mainly contributes to the observed flux in our cases.
From Equations~\eqref{eq.dadmu1} and \eqref{eq.dRdmu}, we find
\begin{align}
    \label{eq.dadmu2}
    \frac{\partial \alpha}{\partial \mu} \sim -\frac{R(\mu - \beta_\sh)}{D\sqrt{1-\mu^2}(1-\mu\beta_\sh)}.
\end{align}
Substituting Equations~\eqref{eq.Jacobian}-\eqref{eq.cos} and \eqref{eq.dadmu2} into Equation~\eqref{eq.Fnu_1}, we find
\begin{align}
    \label{eq.Fnu_2}
    F_\nu (T)&= \frac{1}{D^2} \int _0^{\theta_\mathrm{j}} \diff \theta \sin \theta \nonumber \\ 
    &\quad \times \left. \int _0^{2\pi}\diff \varphi \frac{R^2|\mu - \beta_\sh|}{1-\mu\beta_\sh}\frac{j_\nu}{\alpha_\nu}(1-e^{-\tau_\nu})\right|_{t=T+\mu R/c}.
\end{align}
We obtain Equation~\eqref{eq.Fnu} by substituting 
\begin{align}
    j_\nu &= \frac{j'_{\nu'}}{\Gamma^2(1-\beta \mu)^2} = \frac{\emisd}{4\pi \Gamma^2(1-\beta \mu)^2},\\
    \alpha _\nu &= \Gamma(1-\beta \mu)\alpha'_{\nu'}
\end{align}
into Equation~\eqref{eq.Fnu_2}, where the emission is assumed to be isotropic in the fluid rest frame.

$\Delta s$ in Equation~\eqref{eq.opacity} is found by taking the optically thin limit, $\tau_\nu \rightarrow 0$, in Equation~\eqref{eq.Fnu_2}:
\begin{align}
    \label{eq.Fnu_3}
    F_\nu (T)&\sim \frac{1}{D^2} \int _0^{\theta_\mathrm{j}} \diff \theta \sin \theta \left. \int _0^{2\pi}\diff \varphi \frac{R^2|\mu - \beta_\sh|}{1-\mu\beta_\sh} j_\nu \Delta s\right|_{t=T+\mu R/c}.
\end{align}
Since Equation~\eqref{eq.Fnu_3} should coincide with Equation~\eqref{eq.Fnu_3_thin}, $\Delta s$ is found to be
\begin{align}
    \label{eq.s}
    \Delta s = \frac{R}{12\Gamma^2|\mu - \beta_\sh|}.
\end{align}

We note that Equations~\eqref{eq.Fnu_2} and \eqref{eq.s} change as follows when $\lambda$ is not neglected in Equation~\eqref{eq.dRdmu_middle}:
\begin{align}
    F_\nu (T)&= \frac{1}{D^2} \int _0^{\theta_\mathrm{j}} \diff \theta \sin \theta \int _0^{2\pi}\diff \varphi \nonumber \\ 
    &\quad \times \left. \frac{R^2|\mu - \beta_\sh - \lambda \beta_\sh (1-\mu^2)|}{1-\mu\beta_\sh}\frac{j_\nu}{\alpha_\nu}(1-e^{-\tau_\nu})\right|_{t=T+\mu R/c},\\
    \Delta s &= \frac{R}{12\Gamma^2|\mu - \beta_\sh - \lambda \beta_\sh (1-\mu^2)|}.
\end{align}

\subsection{Interpretation of $\Delta s$ given by Equation~\eqref{eq.s}}\label{app.s}
Figure~\ref{fig.geometry2} gives a visual explanation of $\Delta s$ given by Equation~\eqref{eq.s}.
The left panel shows $\Delta s$ for $\mu > \beta_\sh$, where photons emitted from the tail of the shocked region propagate into the shocked matter and escape through the shock front.
On the other hand, the right panel shows $\Delta s$ for $\mu < \beta_\sh$, where photons emitted from the shock front propagate into the shocked matter and escape through the back side. We below deduce the above interpretation.

Let $(r_1, \mu_1)$ and $t_1$ be the coordinates of an edge on $\Delta s$ and the laboratory time when photons are emitted from that position, respectively.
Similarly, let $(r_2, \mu_2)$ and $t_2$ be the coordinates of the other edge on $\Delta s$ and the time when photons are emitted, respectively.
We note that $r_1$ and $r_2$ correspond to either the radius of the shock front or the tail of the shocked region, depending on the situation.
Since these photons arrive at the observer at $T$, $r_{1,2}$, $\mu_{1,2}$, and $t_{1,2}$ satisfy
\begin{align}
    \label{eq.equivt_2}
    T = t_1 - \frac{\mu_1 r_1}{c} = t_2 - \frac{\mu_2 r_2}{c}.
\end{align}
By construction, $\Delta s$ is given by
\begin{align}
    \Delta s = c(t_1 - t_2).
\end{align}
Since $\alpha$ is constant on a ray, we find from Equation~\eqref{eq.sin_thin} that
\begin{align}
    \label{eq.sameray}
    r_1 \sqrt{1 - \mu_1^2} = r_2 \sqrt{1 - \mu_2^2}.
\end{align}
From Equations~\eqref{eq.equivt_2}-\eqref{eq.sameray}, we find
\begin{align}
    \label{eq.s_equation}
    \frac{\Delta s}{r_2} = \sqrt{\left(\frac{r_1}{r_2}\right)^2 - (1-\mu_2^2)} - \mu_2.
\end{align}
We solve Equation~\eqref{eq.s_equation} below in the cases of $\mu_2 > \beta_\sh$ and $\mu_2 < \beta_\sh$, respectively.

When $\mu_2 > \beta_\sh$, we find that it is relevant to choose
\begin{align}
    \label{eq.r_1_mu2>beta_sh}
    r_1 &= R(t_1) ,\\
    \label{eq.r_2_mu2>beta_sh}
    r_2 &= R(t_2) - w(t_2),
\end{align}
where we omitted the dependence of $R$ on $\theta$, since the jet structure was neglected ($|\lambda| \lesssim 1$) in Equation~\eqref{eq.s}.
That is, photons emitted from the back side of the shocked region propagate into the shocked matter and then escape from the shocked region by passing through the shock front.
Substituting Equations~\eqref{eq.relationbtwR1andR2_thin}, \eqref{eq.w}, \eqref{eq.r_1_mu2>beta_sh} and \eqref{eq.r_2_mu2>beta_sh} into Equation~\eqref{eq.s_equation}, we find
\begin{align}
    \label{eq.quadratic_2}
    \frac{1}{\Gamma_\sh^2}\left(\frac{\Delta s}{R(t_2)}\right)^2
    +2\left[\left(1-\frac{1}{12\Gamma^2}\right)\mu_2 - \beta_\sh\right]\frac{\Delta s}{R(t_2)}
    -\frac{2}{12\Gamma^2}\sim 0,
\end{align}
where we omitted higher order terms proportional to $\Gamma^{-4}$. Solving Equation~\eqref{eq.quadratic_2}, we obtain
\begin{align}
    \label{eq.s_mu2>beta_sh}
    \Delta s \sim \frac{R(t_2)}{12\Gamma^2(\mu_2 - \beta_\sh)},
\end{align}
for $\mu_2 - \beta_\sh \gg O(\Gamma^{-2})$.\footnote{The other solution of the quadratic equation, Equation~\eqref{eq.quadratic_2}, gives $\Delta s < 0$ and is irrelevant.}
Equation~\eqref{eq.s_mu2>beta_sh} is the same as Equation~\eqref{eq.s} for $\mu > \beta_\sh$, if we formally replace $\mu$ and $R$ in Equation~\eqref{eq.s} by $\mu_2$ and $R(t_2)$, respectively.

When $\mu_2 < \beta_\sh$, on the other hand, it is relevant to set
\begin{align}
    \label{eq.r_1_mu2<beta_sh}
    r_1 &= R(t_1) - w(t_1),\\
    \label{eq.r_2_mu2<beta_sh}
    r_2 &= R(t_2),
\end{align}
That is, photons emitted from the shock front propagate into the shocked matter and then escape from the shocked region by passing through the back side.
Substituting Equations~\eqref{eq.relationbtwR1andR2_thin}, \eqref{eq.w}, \eqref{eq.r_1_mu2<beta_sh} and \eqref{eq.r_2_mu2<beta_sh} into Equation~\eqref{eq.s_equation}, we obtain the following equation:
\begin{align}
    \label{eq.quadratic_1}
    &\left[1-\left(1-\frac{2}{12\Gamma^2}\right)\beta_\sh^2\right]\left(\frac{\Delta s}{R(t_2)}\right)^2
    -2\left[\left(1-\frac{2}{12\Gamma^2}\right)\beta_\sh - \mu_2\right]\frac{\Delta s}{R(t_2)}\nonumber \\
    &\quad +\frac{2}{12\Gamma^2} \sim 0,
\end{align}
where we omitted higher order terms proportional to $\Gamma^{-4}$. 
Solving Equation~\eqref{eq.quadratic_1}, we obtain
\begin{align}
    \label{eq.s_mu2<beta_sh}
    \Delta s \sim \frac{R(t_2)}{12 \Gamma^2(\beta_\sh - \mu_2)},
\end{align}
for $\beta_\sh - \mu_2 \gg O(\Gamma^{-2})$.\footnote{The other solution of the quadratic equation, Equation~\eqref{eq.quadratic_1}, gives $\Delta s < 0$ and is irrelevant.}
Equation~\eqref{eq.s_mu2<beta_sh} is the same as Equation~\eqref{eq.s} for $\mu < \beta_\sh$, if we formally put $\mu = \mu_2$ and $R=R(t_2)$ in Equation~\eqref{eq.s}.

We note that the emission from $\mu < \beta_\sh$ is quantitatively important for calculating observed fluxes, although it is suppressed by relativistic beaming, as shown in appendix~C in \citetalias{TI20}.

Lastly, we comment on the case with $\mu = \beta_\sh$, for which $\Delta s$ given by Equation~\eqref{eq.s} diverges.
The divergence of $\Delta s$ is obviously unphysical, but it does not practically matter in our basic equation, Equation~\eqref{eq.Fnu_2}, since the integrand in Equation~\eqref{eq.Fnu_2} does not diverge for $\mu=\beta_\sh$: the integrand $\propto |\mu -\beta_\sh |[1 - \exp(-\tau _\nu)]= 0$ for $\mu = \beta_\sh$.

\section{The absorption coefficient of synchrotron self-absorption} \label{app.ssa}
We derive the absorption coefficient of synchrotron self-absorption, $\alpha'_{\nu'}$, in Equation~\eqref{eq.Fnu}, which is given by Equation~\eqref{eq.alpha_fin} below.

The absorption coefficient of synchrotron self-absorption by electrons with a power-law energy distribution $N'_\mathrm{e}(E'_\mathrm{e})$ is given by \citep{RL}
\begin{align}
    \label{eq.alpha}
    \hat{\alpha}' _{\nu'} = \frac{(p+2)c^2}{8\pi \nu^{\prime 2}} \int _{\gamma '_\mathrm{min}}^\infty \diff \gamma_\mathrm{e}' P'_{\nu'}(\gamma_\mathrm{e}',\psi')\frac{N_\mathrm{e}'(E_\mathrm{e}')}{\gamma_\mathrm{e}'},
\end{align}
where $P'_{\nu'}(\gamma_\mathrm{e}')$ is the synchrotron power per frequency emitted by an electron with a pitch angle $\psi'$:
\begin{align}
    \label{eq.P'}
    P'_{\nu'}(\gamma_\mathrm{e}',\psi') &= \frac{\sqrt{3}\qe^3B'\sin \psi'}{\me c^2}F(x),
\end{align}
with
\begin{align}
    \label{eq.F}
    F(x) &= x\int_x^\infty K_{5/3}(\xi) \diff \xi,\\
    \label{eq.x}
    x &= \frac{4\pi \me c\nu'}{3\qe \gamma_\mathrm{e}^{\prime 2} B'\sin \psi'}.
\end{align}
Substituting Equations~(\ref{eq.e-dist}), (\ref{eq.C}) [for the slow cooling phase, or (\ref{eq.e-dist2}), (\ref{eq.C2}) for the fast cooling phase] and (\ref{eq.P'})-(\ref{eq.x}) to Equation~(\ref{eq.alpha}), we obtain
\begin{align}
    \label{eq.alpha_2}
    \hat{\alpha}' _{\nu'} &= \frac{2^{\tilde{p}+4}(\tilde{p}+2)(\tilde{p}-1)(\sin \psi')^{(\tilde{p}+2)/2}\qe n_\mathrm{R}^\prime}{3^{3/2}\pi^{(\tilde{p}+2)/2}\gamma_\mathrm{min}^{\prime 5}B'}\left(\frac{\nu'}{\nu'_\mathrm{min}}\right)^{-(\tilde{p}+4)/2} \nonumber \\
    &\qquad \times \int_0^{x_\mathrm{M}}x^{(\tilde{p}-2)/2}F(x) \diff x, \\
    x_\mathrm{M} &= \frac{4\pi \me c\nu'}{3\qe \gamma_\mathrm{min}^{\prime 2} B'\sin \psi'}=\frac{\pi \nu'}{4 \nu'_\mathrm{min}\sin \psi'},
\end{align}
where $\gamma_\mathrm{min}^\prime = \min(\gamma_\mathrm{c}^\prime, \gamma_\mathrm{m}^\prime)$, $\nu_\mathrm{min}^\prime = \min(\nu_\mathrm{c}^\prime, \nu_\mathrm{m}^\prime)$, and $\tilde{p}$ stands for $p$ given by Equation~\eqref{eq.p} in the slow cooling phase ($\gamma_\mathrm{m}^\prime < \gamma_\mathrm{c}^\prime$) while $\tilde{p}=2$ in the fast cooling phase ($\gamma_\mathrm{m}^\prime \ge \gamma_\mathrm{c}^\prime$).
Here, we take the average of the pitch angle $\psi$ over all solid angles, while assuming an isotropic distribution of the pitch angle.
However, it is difficult to analytically take the pitch-angle average of Equation~(\ref{eq.alpha_2}) in general cases. Hence, we consider two limiting cases, $\nu'\ll \nu'_\mathrm{min}$ and $\nu'\gg \nu'_\mathrm{min}$, below and connect them at the intermediate regime by interpolation.

When $\nu' \ll \nu'_\mathrm{min}$ (or, equivalently, $x_\mathrm{M} \ll 1$), $F(x)$ is asymptotically given by
\begin{align}
    \label{eq.F_asympt_small}
    F(x) \sim \frac{4\pi}{\sqrt{3}\tilde{\Gamma}\left(\frac{1}{3}\right)}\left(\frac{x}{2}\right)^{1/3} \quad (x \ll 1),
\end{align}
where $\tilde{\Gamma}$ is the Gamma function.
Then, the integration in Equation~(\ref{eq.alpha_2}) is analytically performed by using Equation~(\ref{eq.F_asympt_small}) and, as a result, $\hat{\alpha}'_{\nu'}$ is proportional to $(\sin \psi')^{2/3}$ \citep{Granot99b}. 
Noting the angle average of $(\sin \psi')^{a}$ $(a>0)$ is given by
\begin{align}
    \langle (\sin \psi')^{a} \rangle &:= \frac{1}{4\pi}\int \diff \Omega (\sin \psi')^{a},\\
    \label{eq.gamma_formula}
    &=\frac{\sqrt{\pi}\tilde{\Gamma}\left(1+\frac{a}{2}\right)}{2\tilde{\Gamma}\left(\frac{3+a}{2}\right)},\\
    &=\frac{\sqrt{\pi}a\tilde{\Gamma}\left(\frac{a}{2}\right)}{2(1+a)\tilde{\Gamma}\left(\frac{1+a}{2}\right)},
\end{align}
we obtain the absorption coefficient $\alpha' _{\nu'} = \langle \hat{\alpha}' _{\nu'} \rangle$ for $\nu' \ll \nu'_\mathrm{min}$ from Equation~(\ref{eq.alpha_2}):
\begin{align}
    \label{eq.alpha_lowfreq}
    \alpha' _{\nu'} &= \frac{2^6\pi^{5/6}(\tilde{p}+2)(\tilde{p}-1)\qe n_\mathrm{R}^\prime}{15(3\tilde{p}+2)\tilde{\Gamma}\left(\frac{5}{6}\right)\gamma_\mathrm{min}^{\prime 5}B'}
    \left(\frac{\nu'}{\nu'_\mathrm{min}}\right)^{-5/3} \quad (\nu' \ll \nu'_\mathrm{min}).
\end{align}
We note that Equation~(\ref{eq.alpha_lowfreq}) recovers equation~(18) in \citet{Granot99b} in the relativistic limit $(\Gamma \gg 1)$.\footnote{\citet{Granot99b} considered the synchrotron self-absorption in a self-similar solution of \citet{BM}. Substituting Equations~(\ref{eq.gmin}), (\ref{eq.B}), (\ref{eq.strong_rela_n}), and (\ref{eq.strong_rela_e}) to Equation~(\ref{eq.alpha_lowfreq}), one finds that Equation~(\ref{eq.alpha_lowfreq}) coincides with equation~(18) in \citet{Granot99b} at $\chi=1$ , where $\chi=1$ corresponds to the shock radius.}

For $\nu' \gg \nu'_\mathrm{min}$ (or, equivalently, $x_\mathrm{M} \gg 1$), on the other hand, we can approximately change the interval of the integral in Equation~(\ref{eq.alpha_2}) from $[0,x_\mathrm{M}]$ to $[0,\infty)$, since $F(x)$ exponentially decreases for $x\gg 1$:
\begin{align}
    F(x) \sim \sqrt{\frac{\pi x}{2}}e^{-x} \quad (x \gg 1).
\end{align}
Then, the integral in Equation~(\ref{eq.alpha_2}) is given by
\begin{align}
    \label{eq.difficult_integral}
    \int _0^\infty x^{(\tilde{p}-2)/2}F(x) \diff x &= \frac{2^{(\tilde{p}+2)/2}}{\tilde{p}+2}\tilde{\Gamma}\left(\frac{\tilde{p}}{4}+\frac{11}{6}\right)\tilde{\Gamma}\left(\frac{\tilde{p}}{4}+\frac{1}{6}\right),
\end{align}
where we used eq.~(6.35a) in \citet{RL}.
Using Equations~\eqref{eq.gamma_formula} and \eqref{eq.difficult_integral},
we obtain the absorption coefficient for $\nu' \gg \nu'_\mathrm{min}$ from Equation~(\ref{eq.alpha_2}):
\begin{align}
    \label{eq.alpha_highfreq}
    \alpha' _{\nu'} &= \frac{2^{(3\tilde{p}+8)/2}(\tilde{p}-1)\tilde{\Gamma}\left(\frac{3}{2}+\frac{\tilde{p}}{4}\right)
    \tilde{\Gamma}\left(\frac{11}{6}+\frac{\tilde{p}}{4}\right)
    \tilde{\Gamma}\left(\frac{1}{6}+\frac{\tilde{p}}{4}\right)
    \qe n_\mathrm{R}^\prime}{3^{3/2}\pi^{(\tilde{p}+1)/2}\tilde{\Gamma}\left(2+\frac{\tilde{p}}{4}\right)\gamma_\mathrm{min}^{\prime 5}B'}\nonumber \\
    &\qquad \times
    \left(\frac{\nu'}{\nu'_\mathrm{min}}\right)^{-(\tilde{p}+4)/2} \quad (\nu' \gg \nu'_\mathrm{min}).
\end{align}

The two asymptotes, Equations~(\ref{eq.alpha_lowfreq}) and (\ref{eq.alpha_highfreq}), intersect at a frequency $\nu'_0$ given by
\begin{align}
    \label{eq.nu0}
    \nu'_0 &= 
    \left[\frac{5(\tilde{3}p+2)\tilde{\Gamma}\left(\frac{3}{2}+\frac{\tilde{p}}{4}\right)\tilde{\Gamma}\left(\frac{11}{6}+\frac{\tilde{p}}{4}\right)\tilde{\Gamma}\left(\frac{1}{6}+\frac{\tilde{p}}{4}\right)\tilde{\Gamma}\left(\frac{5}{6}\right)}{(\tilde{p}+2)\tilde{\Gamma}\left(2+\frac{\tilde{p}}{4}\right)}\right]^{6/(3\tilde{p}+2)}
    \nonumber \\
    &\qquad \times \left[\frac{2^{3(3\tilde{p}-4)}}{27\pi^{3\tilde{p}+8}}\right]^{1/(3\tilde{p}+2)} \nu_\mathrm{min}'.
\end{align}
We note that the numerical coefficient in Equation~(\ref{eq.nu0}) is about unity for $2\le \tilde{p}\le 2.22$: $\nu'_0 = 1.07 \nu_\mathrm{min}'$ for $\tilde{p}=2$ and $\nu'_0 = 1.12 \nu_\mathrm{min}'$ for $\tilde{p}=2.22$, for example.
We define the absorption coefficient of synchrotron self-absorption by connecting the two limiting cases as
\begin{align}
    \label{eq.alpha_fin}
    \alpha'_{\nu'} = \alpha'_{\nu_0'} \left(\frac{\nu '}{\nu_0'}\right)^\eta, 
    \quad \eta = \begin{cases}
    -\frac{5}{3} & (\nu' < \nu'_0)\\
    -\frac{\tilde{p}+4}{2} & (\nu' \ge \nu'_0)
    \end{cases}.
\end{align}
The coefficient in Equation~(\ref{eq.alpha_fin}) is defined by
\begin{align}
    \label{eq.alpha0}
    \alpha'_{\nu_0'}=\alpha'_{\nu'}(\nu_0'),
\end{align}
where the right hand side is calculated by using Equation~(\ref{eq.alpha_lowfreq}) or, equivalently, Equation~(\ref{eq.alpha_highfreq}).
We note that \citet{Eerten10} employed $\alpha'_{\nu'}$ with a broken power law bent at $\nu'=\nu'_\mathrm{min}$ with the same exponents as ours but with a different coefficient [See their equation~(A21)].\footnote{Equation~(\ref{eq.alpha_fin}) gives larger $\alpha'_{\nu'}$ than that in \citet{Eerten10} by a factor of $\sim2$ for $2\le p \le 2.22$ if compared for a fixed $\nu'/\nu'_\mathrm{min}$.}


\bsp	
\label{lastpage}
\end{document}